\documentclass[onecolumn,preprintnumbers,amsmath,amssymb,nofootinbib,pagenumbers,unsortedaddress,aps]{revtex4-1}

\usepackage{mathrsfs}
\usepackage{slashed}
\usepackage{bm}

\newcommand{\Tr}{\mathop{\mathrm{Tr}}}
\DeclareMathOperator{\sgn}{sgn}

\usepackage[colorinlistoftodos]{todonotes}
\def\Journal#1#2#3#4{{#1} \textbf{#2}, {\em #3}, #4}

\def\NIMA{\em Nucl. Instrum. Methods A}

\def\PLB{\em Phys. Lett.  B}
\def\PRL{\em Phys. Rev. Lett.}
\def\PRD{\em Phys. Rev. D}

\def\APJ{\em Astrophys. J.}

\def\s{{\,\rm s}}
\def\g{{\,\rm g}}
\def\eV{\,{\rm eV}}
\def\keV{\,{\rm keV}}
\def\MeV{\,{\rm MeV}}

\def\TeV{\,{\rm TeV}}
\def\sv{\left<\sigma v\right>}
\def\({\left(}
\def\){\right)}
\def\cm{{\,\rm cm}}

\def\beq{\begin{equation}}
\def\eeq{\end{equation}}
\def\bear{\begin{eqnarray}}
\def\ear{\end{eqnarray}}

\usepackage{microtype}

\usepackage{hyperref}

\begin{document}

\title{Hadronic and Hadron-Like Physics of Dark Matter}

\author{Vitaly Beylin}
\email{vitbeylin@gmail.com}
\affiliation{%
Research Institute of Physics,
Southern Federal University, 
Prospekt Stachki, 194, 344090 Rostov-na-Donu, Russia
}
\author{Maxim Yu.~Khlopov}
\email{maxim51khl@yahoo.com}
\affiliation{%
Research Institute of Physics,
Southern Federal University,
Prospekt Stachki, 194, 344090 Rostov-na-Donu, Russia
}
\affiliation{%
APC Laboratory 10, rue Alice Domon et L\'eonie Duquet, CEDEX 13, 75205 Paris, France
}
\affiliation{%
National Research Nuclear University ``MEPHI'' (Moscow State Engineering Physics Institute), 31 Kashirskoe Chaussee, Moscow 115409, Russia
}
\author{Vladimir Kuksa}
\email{vkuksa47@mail.ru}
\affiliation{%
Research Institute of Physics,
Southern Federal University,
Prospekt Stachki, 194, 344090 Rostov-na-Donu, Russia
}
\author{Nikolay Volchanskiy}
\email{nikolay.volchanskiy@gmail.com}
\affiliation{%
Research Institute of Physics,
Southern Federal University,
Prospekt Stachki, 194, 344090 Rostov-na-Donu, Russia
}
\affiliation{%
Bogoliubov Laboratory of Theoretical Physics, Joint Institute for Nuclear Research,
Joliot-Curie 6, Dubna 141980, Russia
}

\begin{abstract}The problems of simple elementary weakly interacting massive particles (WIMPs) appeal to extend the physical basis for nonbaryonic dark matter. Such extension involves more sophisticated dark matter candidates from physics beyond the Standard Model (BSM) of elementary particles. We discuss several models of dark matter, predicting new colored, hyper-colored or techni-colored particles and their accelerator and non-accelerator probes. The nontrivial properties of the proposed dark matter candidates can shed new light on the dark matter physics. They  provide interesting solutions for the puzzles of direct and indirect dark matter search.
\end{abstract}


\maketitle

%====================================================================================
%====================================================================================
%====================================================================================

\section{Introduction}

The nature of dark matter is inevitably linked to beyond the Standard Model (BSM) physics of elementary particles. In the lack of direct experimental evidences for this physics, methods of cosmo--particle physics~\cite{ADS,MKH,book,Khlopov:2004jb,bled,newBook} are needed for its study involving proper combination of cosmological, astrophysical and experimental physical~probes. 

The most popular simplest dark matter candidate---elementary weakly interacting massive particles (WIMPs)---finds support neither in direct dark matter searches, nor in searches for supersymmetric (SUSY) particles at the Large Hadron Collider (LHC). The latter removes strong theoretical motivation for WIMPs as the lightest supersymmetric particles and opens the room for wider class of BSM models and corresponding dark matter candidates~\cite{DMRev,bertone,LSSFW,Gelmini,Aprile:2009zzd,Feng:2010gw}.    
 
In this paper, we turn to a possibility of hadronic, hyperhadronic, and composite dark matter candidates. In particular, in the scenario with hadronic dark matter it is suggested that such candidates consist of a new heavy quark and a light standard one. 

It can be shown that the effects of new physics are related to the new massive stable quarks, which fit into the limits imposed by both the electroweak and cosmo--particle data. The bound states of these heavy fermions with light QCD quarks can be considered as (pseudoscalar) neutral dark matter candidates.  Masses of these particles as the lifetime of the charged component are estimated. Besides, we study a low-energy asymptotics of the potential for their interactions with nucleons and with each other. Remind, there is well-known Sommerfeld--Gamov--Sakharov enhancement when heavy particles annihilate. Here, this effect for the states considered is also discussed.

An extension of the Standard Model (SM) with an additional symplectic hypercolor gauge group is analyzed. The extension keeps the Higgs boson of SM as a fundamental field but permits the Higgs to participate in mixing with composite hyperhadrons. New heavy hyperquarks are assumed to be vector-like in the character of their interactions with the intermediate vector bosons. We also consider the properties of pseudo-Nambu--Goldstone (pNG) bosons emerging as a result of dynamical symmetry breaking $\text{SU}(2n_F)\to \text{Sp}(2n_F)$, with $n_F$ being a number of hyperquark flavors. Some versions of the model are invariant under specific global symmetries that ensure the stability of a neutral pseudoscalar field and scalar diquark states (hyperbaryons).  Possible signals of the emergence of these lightest states at colliders are also discussed on the basis of the dark matter (DM) two-component model. Consideration of the DM relic density kinetics allows us to evaluate masses of these neutral stable states and, consequently, to analyze some processes with their participation. Here, we briefly describe  possible channels of the hyperhadron production at colliders having a specific signature of final state. Moreover, there occurs an interesting feature of cosmic ray scattering off the dark matter particles; the~study of a diffuse photon spectrum produced by annihilation of the DM candidates results in a possible prominent manifestation of the two-component structure of dark matter. It is especially important signal because one of the DM components does not interact with vector bosons directly and is, in some sense, invisible in electroweak processes. The generalization of vector-like model symmetry to include three hyperquark flavors significantly expands the spectrum of states, leading to new additional stable hadrons. They can be observed and identified as specific features of the energy and angular spectra of photons and/or leptons recorded mainly by space telescopes. Virtually all additional new hadronic states are quite massive, which prevents their production and study~at~colliders.

The dark atom scenario assumes existence of stable multiple charged particles that can be predicted in some non-supersymmetric BSM models. It involves minimal number of parameters of new physics---the mass of the new charged particles. Particles with charge
~$-2$ bind with primordial helium nuclei in a  neutral $O$He atom.  The nuclear interactions of its $\alpha$-particle shell dominantly determine  cosmological evolution and astrophysical effects of these atoms. However, the nontrivial structure of the $O$He atom with the radius of the Bohr orbit equal to the size of helium nucleus and strongly interacting atomic shell make impossible to apply usual approximations of atomic physics to its analysis. Qualitatively this approach can shed light on the puzzles of direct dark matter searches. It~can give explanation to the observed excess in radiation in positronium annihilation line in the galactic bulge. It can explain the excess of high energy positrons in cosmic rays as indirect effects of composite dark matter. Stable multiple charged constituents of dark atoms are a challenge for their direct search at the LHC and the such searches acquire the meaning of the direct probe of dark atom cosmology.

The paper is organized as follows. In the framework of the hadronic scenario we consider new stable hadrons, in which a new stable heavy quark is bound by the standard QCD interaction with ordinary light quarks (Section \ref{SSQ}). In hyperhadronic models, new heavy quarks are bound by hypercolor strong interactions (Section \ref{hyper}). We also consider the scenario with dark atoms, in which the ordinary Coulomb interaction binds new stable 
~$-2$ charged particle with primordial helium nucleus (Section \ref{darkatoms}). We discuss the  physical motivation for these extensions of the Standard model and their experimental and observational signatures. We conclude (Section \ref{Discussion}) by the discussion of the cross-disciplinary test of these BSM models in the context of cosmo--particle physics.

\section{New Stable Hadrons}\label{SSQ}

In this Section, we consider theoretical and experimental motivations for hadronic dark matter. As a rule, DM candidates are interpreted as stable heavy particles which interact with standard particles through the weak vector bosons (WIMP). The last rigid experimental restrictions on the value of WIMP-nucleon interaction cross section~\cite{3} expel some variants of WIMPs as candidates for dark matter. Therefore, alternative variants are considered in literature, namely the model with fermions from fourth generation, hypercolor models, dark atoms in composite DM, and so on (see~the review~\cite{4} and references therein). It was shown in refs.~\cite{5,6,7,8,9,9a,9b,10} that the existence of hadronic DM candidates, which consist of a new heavy stable quark and a light standard one, is not excluded by cosmological data. In particular, such possibility was carefully considered in the chiral-symmetric extension of SM~\cite{10}.

We consider the scenarios where the strong interaction of new heavy quarks with light standard quarks, which is described by $\text{SU}_\text{C}(3)$ symmetry, forms new stable heavy states. This possibility was analyzed in the extensions of SM with fourth generation~\cite{5,6,7,8,9}, in the framework of mirror and chiral-symmetric models~\cite{10,16}, and in the models with a singlet quark~\cite{11,12,13,14,15,15a}. The simplest variant of the chiral-symmetric model was realized in ref.~\cite{10}, where quark content and quantum numbers of new heavy mesons and fermions were represented, and the low-energy phenomenology of new heavy pseudoscalar mesons was described. It is shown in this work that the existence of new hadrons does not contradict to cosmo--chemical data and precision restrictions on new electroweak effects. Here, we should note that the chiral-symmetric scenario may encounter experimental and theoretical difficulties, which was not analyzed in Ref.~\cite{10}. The scenario with fourth generation and its phenomenology was also considered in literature, although there are strong restrictions from invisible $Z$-decay channel, unitarity of quark-mixing matrix, flavor-changing neutral currents, and others. The principal problem of the extension with fourth generation is contained in new quarks contributions into the Higgs decay channels~\cite{10.9}. It was shown that new quarks contribution to vector gauge boson coupling can be compensated by heavy (with the mass around 50 GeV) neutrino contribution~\cite{5,Ilyin,Novikov}. Then Higgs should have the dominant channel of decay to the fourth neutrino, which is excluded by experimental measurements at the LHC. The proposed solution is that the fourth family gets masses from additional heavy Higgses and the standard Higgs (125 GeV) has suppressed couplings to the fourth family. In this section, we consider the hypothesis of hadronic DM candidates which can be built in the framework of the chiral-symmetrical and singlet quark extensions of SM.

\subsection{Gauge Structure of Chiral-Symmetrical Model with New Quarks}

The chiral-symmetrical extension of the standard set of fermionic fields was considered in ref.~\cite{10}, where the phenomenology of new heavy hadrons is described. In this subsection, we consider the group structure of new quark sector and analyze corresponding gauge interactions of quarks with vector bosons. This aspect was not analyzed in ref.~\cite{10} but has significance in electroweak precision test of the model. In the model under consideration, new multiplets of the up and down quarks has chiral-symmetric structure with respect to the standard set of quarks:
\begin{equation}\label{1.1}
Q=\{Q_R=\begin{pmatrix}
  U\\
  D
  \end{pmatrix}_R\,;
  \,\,\,U_L,\,\,\,D_L\} .
\end{equation}

Thus, in contrast to the Standard Model structure, the right-hand components of new quarks are doublets and left-hand ones are singlets. The structure of covariant derivatives follows from
this definition in the standard way:
\begin{equation}
\begin{aligned}\label{1.2}
D_{\mu}Q_R=&(\partial_{\mu}-ig_1 Y_Q V_{\mu}-\dfrac{ig_2}{2} \tau_a V^a_{\mu}-ig_3 t_iG^i_{\mu}) Q_R;\\
D_{\mu}U_L=&(\partial_{\mu}-ig_1 Y_U V_{\mu}-ig_3 t_iG^i_{\mu}) U_L,\\D_{\mu}D_L=&(\partial_{\mu}-ig_1 Y_D V_{\mu}-ig_3 t_iG^i_{\mu}) D_L.
\end{aligned}
\end{equation}

In Equation (\ref{1.2}), the values $Y_A$, $A=Q$, $U$, $D$, are the hypercharges of quark multiplets (doublets and singlets), $t_i$ are generators of $\text{SU}_\text{C}(3)$ group, which describes the standard color (strong) interaction. We should note that the coupling constants $g_1$ and $g_2$ are equal to the standard ones at the energy scale, where the chiral symmetry is restored. At low energy they depend on the details of symmetry violation scenario. Here, the gauge boson fields $V^a_{\mu}$ are chiral partner of the standard gauge bosons which are expected to be superheavy. The status of the abelian gauge field $V_{\mu}$ depends on its physical interpretation in low-energy electromagnetic processes. If we interpret abelian gauge field $V_{\mu}$ as standard one, then the mixing of $V_{\mu}$ and $V^3_{\mu}$ in the standard way is forbidden, because it leads to contradictions with precision electroweak measurements. Moreover, a direct interpretation of the field $V_{\mu}$ and the weak hypercharge $Y_Q=\bar{q}$, where $\bar{q}$ is an average electric charge of a quark multiplet (in~our case, doublet and singlets), leads to wrong a $V-A$ structure
of photon interaction with fermions. To escape these obstacles, we have two options:
\begin{itemize}
\item to interpret the field $V_{\mu}$ as a new non-standard abelian field which mixes
with $V^3_{\mu}$ in analogy with the standard procedure,
\item to assume that $V_{\mu}$ is standard abelian field which does not mix with $V^3_{\mu}$.
\end{itemize}

The first option leads to a mirror world with exotic ``electroweak'' vector bosons (in particular, mirror or dark photons) and standard QCD-like strong interactions. In the framework of this scenario, it is difficult to build coupled heavy-light states of type $(qQ)$ and satisfy cosmo--chemical restrictions which were considered in ref.~\cite{10}. So, we consider the second option and analyze the interaction of $V_{\mu}$ with additional heavy quark sector. This scenario needs a redefinition of the hypercharge operator $\hat{Y}_Q$, which we consider below. The standard definition of the hypercharge operator $\hat{Y}_Q=(\hat{q}-\hat{t})$ in the case of the doublet $Q$ can be realized with the help of $2\times 2$ matrices of charge $\hat{q}$ and isospin $\hat{t}=\tau_3/2$. This operator in matrix representation acts on the standard left quark doublet as follows:
\begin{equation}\label{1.3}
  \hat{Y}_Q Q_L=
  \begin{pmatrix}
  \hat{q}_u-1/2&0\\
  0&\hat{q}_d+1/2
  \end{pmatrix}
  \begin{pmatrix}
  u\\
  d
  \end{pmatrix}_L.
\end{equation}

The operator of charge $\hat{q}$ in (\ref{1.3}) is defined by equalities $\hat{q}_u u=2/3\,u$ and $\hat{q}_d d=-1/3\,d$. So, it follows from (\ref{1.3}) that
\begin{equation}\label{1.4}
  \hat{Y}_Q Q_L=\dfrac{1}{6}
  \begin{pmatrix}
  1&0\\
  0&1
  \end{pmatrix}
  \begin{pmatrix}
  u\\
  d
  \end{pmatrix}_L=\dfrac{1}{6}\cdot Q_L.
\end{equation}

Thus, the operator action reduces to multiplication by the coefficient $1/6$ (one half of the average charge of the doublet) and the standard $V$-structure of $\gamma$- and $V-A$ structure of $Z$-interactions arise as a result of the mixing of $A_{\mu}$ and $W^3_{\mu}$. From this simple analysis, it follows that the presence of the isospin operator in definition of the hypercharge operator is connected with the presence of singlet-triplet mixing. So, the hypercharge operator in the absence of mixing has the form $\hat{Y}_Q=\hat{q}$ (without $\hat{t}$):
\begin{equation}\label{1.4a}
  \hat{Y}_Q Q_R=
  \begin{pmatrix}
  \hat{q}_u&0\\
  0&\hat{q}_d
  \end{pmatrix}
  \begin{pmatrix}
  U\\
  D
  \end{pmatrix}_R.
\end{equation}

Taking into account equalities $\hat{q}_u\cdot U_R=\hat{Y}_U\cdot U_L =2/3\cdot U$ and $\hat{q}_d\cdot D_R=\hat{Y}_D\cdot D_L =-1/3\cdot D$,  it~follows from the Equation~(\ref{1.2}) that vector-like interactions
of physical fields $\gamma$ and $Z$ with the new quarks $U=U_R+U_L$ and $D=D_R+D_L$:
\begin{equation}\label{1.5}
\mathscr{L}_Q^{int}=g_1 V_{\mu}\bar{Q}\gamma^{\mu}\hat{q} Q = g_1 (c_w A_{\mu}- s_w Z_{\mu}) (q_U\bar{U} \gamma^{\mu} U - q_D\bar{D} \gamma^{\mu} D),
\end{equation}
where we omit the strong interaction term which has the standard structure. In the expression (\ref{1.5}), abelian field $V_{\mu}$ is the standard mixture of the physical fields $A_{\mu}$ and $Z_{\mu}$, $c_w=\cos\theta_w$, $s_w=\sin\theta_w$, $g_1 c_w=e$ and $\theta_w$ is the Weinberg angle. Indirect limits for the new quarks follow from electroweak measurements of FCNC 
processes and the value of polarizations. Since the new fermions are stable, there are no FCNC processes in our scenario. The constraints which are caused by the vector-boson polarization measurements will be considered below.

\newpage
\subsection{The Extension of Standard Quark Sector with Stable Singlet Quark}

There are many scenarios of SM extensions with singlet (isosinglet) quarks which are considered in literature. The singlet quark (SQ) is usually defined as a Dirac fermion with the quark quantum numbers having the standard $\text{U}_Y(1)$ and $\text{SU}_\text{C}(3)$ gauge interactions. In contrast to a standard quark, it~is a singlet with respect to $\text{SU}_\text{W}(2)$ transformations, that is, it does not interact with the non-abelian weak charged boson $W$. The high-energy origin and low-energy phenomenology of SQ were discussed in literature (see the recent works~\cite{9a,Eberhardt,Botella,Kumar}, and references therein). As a rule, SQ is supposed to be an unstable particle, which is caused by the mixing of SQ with ordinary quarks. Such mixing leads to FCNC appearing at the tree level, which is absent in SM. This results in an additional contributions to rare lepton decays and $M^0-\bar{M}^0$ oscillations. There are strong restrictions on the value of singlet-ordinary quark mixing. Here, we suggest an alternative variant with a stable SQ, namely, the scenario with the absence of such mixing. Further, we analyze this variant and apply it to the description of possible DM candidate. Because the SQ together with ordinary quarks are in confinement they form the bound states of type $(Sq)$, $(Sqq)$, $(SSq)$, and more complicated. Here, we consider the main properties of $(Sq)$-states and describe the lightest state $M^0=(\bar{S}q)$ (which should be stable) as DM particle.

Further, we analyze the scenario with SQ, which in general case can be up, $U$, or down, $D$, type ($q=2/3$ or $q=-1/3$ respectively). According to the above definitions, the minimal Lagrangian describing interactions of the singlet quark $S$ with the gauge bosons is as follows:
\begin{equation}\label{2.1}
\mathscr{L}_S=i\bar{S}\gamma^{\mu}(\partial_{\mu}-ig_1 q V_{\mu} -ig_s t_a G^a_{\mu})S - M_S \bar{S} S.
\end{equation}

In (\ref{2.1}), the hypercharge $Y/2=q$ is a charge of the $S$, $t_a=\lambda_a/2$ are generators of $\text{SU}_\text{C}(3)$ group, and $M_S$ is a phenomenological mass of the $S$. It can
not get mass by the Higgs mechanism because the corresponding term is forbidden by SU(2) symmetry. However, the mass term in expression (\ref{2.1}) is allowed by the symmetry of the model. The abelian part in (\ref{2.1}) contains the interactions of SQ with photon and $Z$-boson:
\begin{equation}\label{2.2}
\mathscr{L}_S^{int}=g_1 q V_{\mu}\bar{S}\gamma^{\mu} S = q g_1 (c_w A_{\mu}- s_w Z_{\mu})\bar{S}\gamma^{\mu} S.
\end{equation}

In expression (\ref{2.2}), the values $c_w=\cos\theta_w$, $s_w=\sin\theta_w$, $g_1 c_w=e$ and $\theta_w$ is the Weinberg angle. We should note that the interaction of SQ with the vector bosons has a vector-like form, so SQ is usually called a vector quark~\cite{Botella,Kumar}.

Now, we take into account the restrictions on the processes with SQ participation, which follow from the experimental data. New sequential quarks are excluded by LHC data on Higgs properties~\cite{Eberhardt}. Because SQ does not interact with the Higgs doublet, it is not excluded by data on Higgs physics. The limits on new quarks for colored factors $n_{eff}=2,3,6$ are about 200 GeV, 300 GeV, and 400~GeV respectively~\cite{Llorente}. As we show further, these limits are much less than our estimations with the assumption that the new heavy quark is a DM particle. The scenario with the long-lived heavy quarks, which takes place when SQ slightly mixes with an ordinary quark, was discussed in the review~\cite{9a}.

\subsection{Constraints on the New Quarks Following from the Precision Electroweak Measurements}

The constraints on the new heavy quarks follow from the electroweak measurements of FCNC and vector boson polarization. As was noted earlier, there is no mixing of the new quarks with the standard ones and the new quarks do not contribute to rare processes. The contributions of the new quarks to polarization tensors of the vector bosons are described by the Peskin--Takeuchi (PT) parameters~\cite{Peskin}. From
Equations~(\ref{1.5}) and (\ref{2.2}), it follows that interactions of the new quarks with the vector bosons in both scenarios (the chiral-symmetrical extension and the model with SQ) have the same
structure and their contributions into the PT parameters can be described by general expressions. In the models under consideration, the new heavy quarks contribute into polarization tensors of
$\gamma$ and $Z$-bosons, namely $\Pi_{\gamma\gamma}$, $\Pi_{\gamma Z}$, $\Pi_{ZZ}$. Note, because the $W$-boson does not interact with the new
quarks, $\Pi_{WW}=0$. To extract the transverse part $\Pi(p^2)$ of the polarization tensor $\Pi_{\mu\nu}(p^2)$, we have used the
definition $\Pi_{\mu\nu}(p^2)=p_{\mu}p_{\nu}P(p^2) + g_{\mu\nu}\Pi(p^2)$. In our case, $\Pi_{ab}(0)=0$, $a$, $b=\gamma$, $Z$ and the PT parameters, which we take from~\cite{Burgess}, can be
represented by the following expressions:
\begin{equation}
\begin{aligned}\label{2.3}
 S=&\dfrac{4s^2_w c^2_w}{\alpha}[\dfrac{\Pi_{ZZ}(M^2_Z,M^2_Q)}{M^2_Z}-\dfrac{c^2_w-s^2_w}{s_w c_w}\Pi^{'}_{\gamma Z}(0,M^2_Q)-\Pi^{'}_{\gamma\gamma}(0,M^2_Q)];
\,\,\,T=-\dfrac{\Pi_{ZZ}(0,M^2_Q)}{\alpha M^2_Z}=0;\\
U=&-\dfrac{4s^2_w}{\alpha} [c^2_w\dfrac{\Pi_{ZZ}(M^2_Z,M^2_Q)}{M^2_Z}+2s_w c_w \Pi^{'}_{\gamma Z}(0,M^2_Q)+s^2_w\Pi^{'}_{\gamma\gamma}(0,M^2_Q)].
\end{aligned}
\end{equation}

In Equation~(\ref{2.3}), $\alpha=e^2/4\pi$, $M_Q$ is a mass of the new heavy quark and $\Pi_{ab}(p^2)$ are defined at $p^2=M^2_Z$ and $p^2=0$. In (\ref{2.3}), the functions $\Pi_{ab}(p^2,M^2_Q)$, $a,b=\gamma,Z$, can be represented in a simple~form:
\begin{equation}
\begin{aligned}\label{2.4}
\Pi_{ab}(p^2,M^2_Q)=&\dfrac{g_1^2}{9\pi^2}k_{ab}F(p^2,M^2_Q);\,\,\,k_{ZZ}=s^2_w,\,k_{\gamma\gamma}=c^2_w,\,k_{\gamma Z}=-s_w c_w;\\
F(p^2,M^2_Q)=&-\dfrac{1}{3}p^2+2M^2_Q+2A_0(M^2_Q)+(p^2+2M^2_Q)B_0(p^2,M^2_Q),
\end{aligned}
\end{equation}
where we take into account the contribution of the new quark with $q=2/3$. The function $F(p^2,M^2_Q)$ in Equation~(\ref{2.4}) contains divergent terms in the one-point, $A_0(M^2_Q)$, and two-point, $B_0(p^2,M^2_Q)$, Veltman functions. These terms are compensated exactly in the physical parameters $S$, $T$, and $U$, defined by the expressions (\ref{2.3}). In the case of the D-type quark, the contributions are four times smaller. Using the standard definitions of the functions $A_0(M^2_Q)$ and $B_0(p^2,M^2_Q)$ and the equality  $B^{'}_0(0,M^2_Q)=M^2_Q/6$, we get:
\begin{equation}\label{2.5}
S=-U=\dfrac{k s^4_w}{9\pi}[-\dfrac{1}{3}+2(1+2\dfrac{M^2_Q}{M^2_Z})(1-\sqrt{\beta}\arctan\dfrac{1}{\sqrt{\beta}})].
\end{equation}

In Equation~(\ref{2.5}), $\beta=4M^2_Q/M^2_Z -1$, $k=16(4)$ at $q=2/3(-1/3)$, and $k=20$ in the case of the chiral-symmetric model. We have checked by direct calculations that in the limit of infinitely heavy masses of the new quarks, $M^2_Q/M^2_Z\to \infty$, the parameters $S$ and $U$ go to zero as $\sim M^2_Z/M^2_Q$. From Equation~(\ref{2.5}) it follows that for the value of mass $M_Q>500$ GeV the parameters $S, U <10^{-2}$. Experimental limits are represented in the review~\cite{PDG}:
\begin{equation}\label{2.6}
S=0.00 +0.11(-0.10),\,\,\,U=0.08\pm 0.11,\,\,\,T=0.02+0.11(-0.12).
\end{equation}

 Thus, the scenarios with the new heavy quarks do not contradict to the experimental electroweak~restrictions.

At the quark-gluon phase of the evolution of the Universe, the new heavy quarks strongly interact with the standard ones. So, there are strong processes of
scattering and annihilation into gluons and quarks, $Q\bar{Q} \to gg$ and $Q\bar{Q}\to q\bar{q}$. Additional contributions to these processes through
electroweak channels $Q\bar{Q} \to \gamma\gamma,\,ZZ$ give small differences for cross sections in the scenarios under consideration. Cross sections of annihilation are derived from the expressions for annihilation of gluons and light quarks into heavy quarks $gg\to Q\bar{Q}$ and $q\bar{q}\to Q\bar{Q}$~\cite{PDG}. The cross section of
two-quark annihilation~\cite{PDG}:
\begin{equation}\label{2.7}
\dfrac{d\sigma}{d\Omega}(\bar{q}q\to\bar{Q}Q)=\dfrac{\alpha_s^2}{9s^3}\sqrt{1-\dfrac{4M^2_Q}{s}}[(M^2_Q-t)^2+(M^2_Q-u)^2+2M^2_Q s],
\end{equation}
where $M_Q$ is a mass of the heavy quark ($M_Q\gg m_q$). At the threshold, $s\approx 4M^2_Q$, the parameters $t\approx u \approx -M^2_Q$. The cross section of the process $\bar{Q}Q\to
\bar{q}q$ can be derived from (\ref{2.7}) by reversing time. In this limit, from (\ref{2.7}), we get
\begin{equation}\label{2.8}
\dfrac{d\sigma}{d\Omega}(\bar{Q}Q\to\bar{q}q)=\dfrac{\alpha_s^2}{18v_r M^2_Q
},
\end{equation}
where $v_r$ is a relative velocity of the heavy quarks. The cross section of two-gluon annihilation~\cite{PDG}:
\begin{equation}\label{2.9}
\dfrac{d\sigma}{d\Omega}(gg\to\bar{Q}Q)=\dfrac{\alpha_s^2}{32s}\sqrt{1-\dfrac{4M^2_Q}{s}}F(s,t,u;M^2_Q),
\end{equation}
where exact expression for the function $F(s,t,u;M^2_Q)$ is rather complicated. In the approximation $s\approx 4M^2_Q$, we get $F\approx 7/6$, and the cross section
of the reversed process $\bar{Q}Q\to gg$ is:
\begin{equation}\label{2.10}
\dfrac{d\sigma}{d\Omega}(\bar{Q}Q\to gg)=\dfrac{7\alpha_s^2}{3\cdot 128 v_r M^2_Q}.
\end{equation}

Now, we give the expressions for the total cross section of two-gluon and two-quark annihilation.  Two-gluon cross section in the low-energy limit:
\begin{equation}\label{2.11}
 \sigma(Q\bar{Q}\to gg)v_r=\dfrac{14\pi}{3}\dfrac{\alpha^2_s}{M^2_Q},
\end{equation}
where $\alpha_s=\alpha_s(M_Q)$. Two-quark cross section in the limit $m_q\to 0$:
\begin{equation}\label{2.12}
\sigma(Q\bar{Q}\to q\bar{q})v_r=\dfrac{2\pi}{9}\dfrac{\alpha^2_s}{M^2_Q}.
\end{equation}

From expressions (\ref{2.11}) and (\ref{2.12}), it follows that the two-gluon channel dominates. The value of cross section of $\sigma v_r \sim 1/M_Q^2$ and the remaining concentration of the heavy component may be dominant at the end of the quark-gluon phase of the evolution. At the hadronization stage, the new heavy quarks, which participate in the standard strong interactions, form coupled states with ordinary quarks. Here, we
consider neutral and charged states of type $(qQ)$. The lightest of them is stable and can be suggested as a DM candidate. Here, we should note that the dominance of the heavy component before the transition from the quark-gluon to hadronization phase may be connected with the dominance of dark
matter relative to ordinary matter. The possibility of the heavy-hadron existence was analyzed in~\cite{10}. It was shown that this possibility does not contradict to cosmo--chemical data. This conclusion was drawn taking account of the repulsive strong interaction of new hadrons with nucleons. This effect will be qualitatively
analyzed in the next section.

The constraints on the new heavy hadrons, which follow from the cosmo--chemical data, will be discussed in the following subsection. Such constraints on the new hadrons as strongly interacting carrier of dark matter (SIMP) follow also from astroparticle physics. The majority of restrictions refer to the mass of the new hadrons or available mass/cross section parameters space
\cite{Starkman,Wolfram,Starkman2,Javorsek,Mitra,Mack,McGuire:2001qj,McGuire2,ZF}. \mbox{As a rule}, the low limits on the mass value do not exceed 1--2 TeV which are an order of magnitude less then our previous estimations of DM particle mass
\cite{15a} (see also the estimations in this work). Here, we should pick out the results of research in~\cite{McGuire:2001qj,McGuire2}, where it was reported that the mass of new
hadrons \mbox{$M\gtrsim 10^2$ TeV}. Further we represent almost the same estimation taking account of Sommerfeld--Gamov--Sakharov enhancement effect in annihilation cross section. An additional and more detailed information on restrictions, which follows from XQC experiments, can be found in the Refs.~\cite{XQC,XQC1}. In the next section, we represent effective theory of low-energy
interaction of new hadrons. From this consideration, it follows that the value of interaction of the new hadrons with nucleons is of the same order of magnitude as the hadronic one (see Ref.~\cite{10}).

\subsection{Composition of New Heavy Hadrons and Long-Distance Interactions with Nucleons}

Due to the strong interaction new quarks together with standard ones form the coupled meson and fermion states, the lightest of which are stable. Classification of such new heavy
hadrons was considered in ref.~\cite{10}, where the main processes with their participation were analyzed. In the Table~\ref{tab1}, we represent the quantum numbers and quark content of new
mesons and fermions for the case of $U$- and $D$-type of new quarks.

\begin{table}
\centering\renewcommand{\arraystretch}{1.9}
\caption{Classification of new hadrons.\label{tab1}}
\begin{ruledtabular}
\begin{tabular}{llll}
$J^P=0^-$         &$T=\dfrac{1}{2}$    &$M_U=(M^0_U\,M^-_U)$   &$M^0_U=\bar{U}u$,\, $M^-_U=\bar{U}d$\\ 
  $J^P=0^-$         &$T=\dfrac{1}{2}$    &$M_D=(M^+_D\,M^0_D)$
&$M^+_D=\bar{D}u$,\, $M^0_D=\bar{D}d$\\ 
 $J=\dfrac{1}{2}$   &$T=1$              &$B_{1U}=(B_{1U}^{++}\,B_{1U}^+\,B_{1U}^0)$ &$B_{1U}^{++}=Uuu,B_{1U}^+=Uud,B_{1U}^0=Udd$\\ 
$J=\dfrac{1}{2}$ &$T=1$
&$B_{1D}=(B_{1D}^+\,B_{1D}^0\,B_{1D}^-)$ &$B_{1D}^{+}=Duu,\,B_{1D}^{-}=Ddd,B_{1D}^0=Dud$\\

$J=\dfrac{1}{2}$    &$T=\dfrac{1}{2}$     &$B_{2U}=(B^{++}_{2U}\,B^+_{2U})$ &$B^{++}_{2U}=UUu,B^+_{2U}=UUd$\\
$J=\dfrac{1}{2}$    &$T=\dfrac{1}{2}$
&$B_{2D}=(B^0_{2D}\,B^-_{2D})$ &$B^0_{2D}=DDu,B^-_{2D}=DDd$\\
$J=\dfrac{3}{2}$ &$T=0$ &$(B^{++}_{3U})$ &$B^{++}_{3U}=UUU$\\ 
$J=\dfrac{3}{2}$ &$T=0$ &$(B^-_{3D})$ &$B^-_{3D}=DDD$\\
\end{tabular}
\end{ruledtabular}
\end{table}

Most of the two- and three-quark states, which are represented in Table~\ref{tab1}, were considered also in Refs.~~\cite{9a,10.8}. Ref.~\cite{13d} considered an alternative for the DM candidates
which are electromagnetically bound states made of terafermions. Here, we propose the neutral $M^0$-particles as candidates for DM.
Another possibility is discussed in Refs.~\cite{7,8,9}---new charged hadrons exist but are hidden from detection. Namely, the particles with charge $q=-2$ are bound with primordial helium. In our case, the interactions of baryons $B_{1Q}$ and $B_{2Q}$, where $Q=U, D$,  are similar to the nucleonic interactions. These particles may compose heavy atomic nuclei
together with nucleons. 

Evolution of new hadrons was qualitatively studied in~\cite{10} and, here, we briefly reproduce this analysis for both cases.
Matter of stars and planets may contain stable $U$-type particles $M^0_U,\, B^+_{1U}$ and $B^{++}_{2U}$ as well as $B^{++}_{3U}$ and $\bar{B}^{++}_{3U}$. The antiparticles
$\bar{M}^0_U,\,\bar{B}^+_{1U}$ and $\bar{B}^{++}_{2U}$ are burning out due to interactions with nucleons $N$:
\begin{equation}\label{3.8}
\bar{M}^0_U+N\to B^+_{1U}+X,\,\,\,\bar{B}^+_{1U}+N\to M^0_U+X,\,\,\,\bar{B}_{1U}^{++}+N\to 2M^0_U+X,
\end{equation}
where $X$ are leptons or photons in the final state. There are no Coulomb barriers for the reactions~(\ref{3.8}), so the particles $\bar{M}^0_U,\,\bar{B}^+_{1U}$ and
$\bar{B}^{++}_{2U}$ burn out during the evolution of the Universe. Other stable particles participate in reactions with annihilation of new quarks:
\begin{equation}\label{3.9}
M^0_U+B_{3U}^{++}\to B_{2U}^{++}+X,\,\,\,M^0_U+B_{2U}^{++}\to B_{1U}^{+}+X,\,\,\,M^0_U+B_{1U}^+\to p+X,\,\,\,\bar{B}_{3U}^{++}+B_{3U}^{++}\to X.
\end{equation}

 The Coulomb barrier may appear in the
last reaction in (\ref{3.9}), however, simple evaluations which are based on the quark model show that the reaction
\begin{equation}\label{3.10}
B_{3U}^{++}+N\to 3\bar{M}^0_U+X
\end{equation}
 is energetically preferred.
Thus, there are no limits for total  burning out of $U$-quarks and, along with them, all positively charged new hadrons to the level compatible with cosmological restrictions. The~rest
of the antiquarks $\bar{U}$ in accordance with (\ref{3.10}) may exist inside neutral $M^0_U$-particles only. Concentration of these particles in matter is determined by the baryon
asymmetry in the new quarks sector.

 In the case of $D$-type hadrons, $M^+_D,\,B^0_{1D},\,B^0_{2D}$ and $B^-_{3D}$ particles and $\bar{B}^-_{3D}$ antiparticles may be situated in matter medium, in the inner part of stars, for example.
 In analogy with (\ref{3.8})
 the reactions for the case of down type hadrons can be represented in the form
\begin{equation}\label{3.11}
\bar{M}^+_D+N\to B^0_{1D}+X,\,\,\,\bar{B}^0_{1D}+N\to M^+_D+X,\,\,\,\bar{B}_{2D}^0+N\to 2M^+_D+X.
\end{equation}

It should be noted that the Coulomb barriers to them are absent. Annihilation of new $D$-quarks goes through the following channels:
\begin{equation}\label{3.12}
M^+_D+B_{3D}^-\to B_{2D}^0+X,\,\,\,M^+_D+B_{2D}^0\to B_{1D}^0+X,\,\,\, M^+_D+B_{1D}^0\to p+X,\,\,\,\bar{B}_{3D}^-+B_{3D}^-\to X.
\end{equation}

From this qualitative analysis and cosmo--chemical restrictions the conclusion was done that the baryon asymmetry in new quark sector exists. This asymmetry has a sign which is opposite to the ordinary baryon
asymmetry sign. This conclusion mainly follows from the ratios
``anomalous/natural'' hydrogen $C\leqslant 10^{-28}$ for $M_Q\lesssim 1\,\mbox{TeV}$~\cite{Smith} and anomalous helium $C\leqslant 10^{-12} - 10^{-17}$ for \mbox{$M_Q\leq 10~\mbox{TeV}$~\cite{Muller}}. In the case of up hadrons, the state $B^+_{1U}=(Uud)$ is heavy proton which can form anomalous hydrogen. The anomalous state $B^+_{1U}$ at hadronization phase
can be formed by coupling of quarks $U, u, d$ and as a result of reaction $\bar{M}^0_U + N \to B^+_{1U} + X$ (the first reaction in (\ref{3.8})). The~antiparticles
$\bar{B}^+_{1U}$ are burning out due to the reaction $\bar{B}^+_{1U} + N \to M^0_U +X$. The states like $(pM^0_U)$ can also manifest itself as anomalous hydrogen. But in~\cite{10} it was shown that the  interaction of $p$ and $M^0_U$ has a potential barrier. So, the formation of the coupled states $(pM^0_U)$ is strongly suppressed. Baryon symmetry of new quarks is not excluded when they are superheavy.

The hadronic interactions are usually described in the meson exchange approach with the help of an effective Lagrangian. Low-energy baryon-meson interaction was described in
\cite{18} by $\text{U}(1)\times \text{SU}(3)$ gauge theory. There, U(1)-interaction corresponds to exchange by singlet vector meson and SU(3) is group of unitary
symmetry. Field contents and structure of Lagrangian, for our case, are represented in~\cite{10}. It was shown that the dominant contribution to this interaction is caused by vector meson exchange. We apply this Lagrangian for analysis of $MN$ and
$MM$ interactions. The part of the Lagrangian which will be used further is as follows:
\begin{equation}
\begin{aligned}\label{4.1}
\mathscr{L}_{int}(V,N,M)&=g_{\omega}\omega^{\mu}\bar{N}\gamma_{\mu}N + g_{\rho}\bar{N}\gamma_{\mu}\tau^a\rho_a^{\mu}N
       +ig_{\omega M}\omega^{\mu}(M^{\dagger}\partial_{\mu}M-\partial_{\mu}M^{\dagger}M)\\
        &+ ig_{\rho M} (M^{\dagger}\tau^a\rho_a^{\mu}\partial_{\mu}M-
       \partial_{\mu}M^{\dagger}\tau^a\rho_a^{\mu}M).
\end{aligned}
\end{equation}

In (\ref{4.1}), $N=(p,n)$ is doublet of nucleons, $M=(M^0,\,M^-),\,M^{\dagger}=(\bar{M}^0,\,M^+)$ are new pseudoscalar mesons. Coupling constants are the following ones~\cite{10}:
\begin{equation}\label{4.2}
g_{\rho}=g_{\rho M}=g/2,\,\,\,g_{\omega}=\sqrt{3}g/2\cos\theta ,\,\,\,g_{\omega M}=g/4\sqrt{3}\cos\theta,\,\,\,
        g^2/4\pi\approx 3.16,\,\,\,\cos\theta =0.644.
\end{equation}

Note, the effective strong interaction does not depend on the type of new heavy quark, so we omit subscriptions $U$ and $D$ in (\ref{4.1}). We should note that  one-pion
exchange is absent because $MM\pi$-vertex is forbidden due to parity conservation.

The potential $V(R)$ and amplitude $f(q)$ in Born approximation are connected by the relation
\begin{equation}\label{4.3}
V(\vec{r})=-\dfrac{1}{4\pi^2\mu}\int f(q)\exp(i\,\vec{q}\, \vec{r})\, d^3q,
\end{equation}
where $\mu$ is a reduced mass and we consider the case of non-polarized particles. The potential of $MN$-interaction was calculated in ref.~\cite{10}, where the
relation $f(q)=-2\pi i\mu F(q)$ was used (here,~$F(Q)$ is Feynman amplitude). It was shown in~\cite{10} that scalar and
two-pion exchanges are strongly suppressed. The potentials of various pairs from  $M=(M^0,M^-)$ and $N=(p,n)$ are described by the following
expressions:
\begin{equation}\label{4.4}
V(M^0,p;r)=V(M^-,n;r)\approx V_{\omega}(r)+V_{\rho}(r),\,\,\, V(M^0,n;r)=V(M^-,p;r)\approx V_{\omega}(r)-V_{\rho}(r).
\end{equation}

In (\ref{4.4}), the terms $V_{\omega}(r)$ and $V_{\rho}(r)$ are as follows:
\begin{equation}\label{4.5}
V_{\omega}(r)=\dfrac{g^2K_{\omega}}{16\pi\cos^2\theta}\,\dfrac{1}{r}\,\exp(-\dfrac{r}{r_{\omega}}),\,\,\,V_{\rho}(r)=\dfrac{g^2K_{\rho}}{16\pi}\,\dfrac{1}{r}\,\exp(-\dfrac{r}{r_{\rho}}),
\end{equation}
where $K_{\omega}=K_{\rho}\approx0.92,\,\,r_{\omega}=1.04/m_{\omega},\,\,r_{\rho}=1.04/m_{\rho}$. Using these values and approximate equality $m_{\omega}\approx m_{\rho}$, we rewrite
the expressions (\ref{4.4}):
\begin{equation}\label{4.6}
V(M^0,p;r)=V(M^-,n;r)\approx 2.5\,\dfrac{1}{r}\,\exp(-\dfrac{r}{r_{\rho}}),\,\,\, V(M^0,n;r)=V(M^-,p;r)\approx 1.0\,\dfrac{1}{r}\,\exp(-\dfrac{r}{r_{\rho}}).
\end{equation}

Two important phenomenological conclusions follow from the expressions (\ref{4.6}). All pairs of particles have repulsive ($V>0$) potential and the existence of a barrier prevents the formation of the coupled states $(pM^0)$, that is, anomalous protons. 
 
The potential of $MM$ interaction can be built in full analogy with $MN$-case. To find the sign of the
potential, we use the following non-relativistic limit:
\begin{equation}\label{4.7}
\mathscr{L}=\mathscr{L}_0+\mathscr{L}_{int}\,\longrightarrow W_k(m,v)-V(r,t),\,\,\,m\dot{v}=-\dfrac{\partial U(r,t)}{\partial r}
\end{equation}
where $W_k(m,v)$ is kinetic term, $V(r,t)$ is potential and we separate the spatial and temporal variables. From this definition and the expression for energy, $E=W +V$, it follows
that at long distance, where the monotonically decreasing function $V(r)$ is positive, $V>0$, this function describes repulsive potential. Here, we use a relation between $\mathscr{L}_{eff}(q)$ and amplitude $F(q)$, namely $F(q)=ik\mathscr{L}_{eff}(q)$, where
$k>0$. Then, we get equality for the sign of $V$ and $F$, $signum(V)=signum(iF)$. Here, the amplitude $F(q)$ is determined by
one-particle exchange diagrams for the process $M_1M_2\to M^{'}_1M^{'}_2$. The vertices are defined by the low-energy Lagrangian (\ref{4.1}). Then we
check that $MN$-interactions have a repulsive character. Note that Lagrangians of $NM^0$ and $N\bar{M}^0$ have opposite signs. This is caused by different signs of vertices $\omega M^0 M^0$ and $\omega \bar{M}^0 \bar{M}^0$, which give dominant contribution. This effect follows from the differential structure of (\ref{4.1})
and operator structure of field function of the $M$-particles. We get the vertices $\omega(q) M^0(p) M^0(p-q)$ and $\omega(q) \bar{M}^0(p) \bar{M}^0(p-q)$ in
momentum representation with opposite signs, $\mathscr{L}_{eff}=\pm g_{\omega M}(2p-q)$, respectively. This leads to the potentials of interactions through $\omega$ exchange, which is repulsive
for the case of $N M$ and attractive for $N \bar{M}$ scattering. Thus, the absence of a potential barrier gives rise to the problem of coupled states $p\bar{M}^0$. To overcome this problem, we assume the existence of asymmetry in the sector of new quarks or that the particles $\bar{M}^0$ are superheavy. Interactions of baryons $B_1$ and $B_2$ are similar to the nucleonic one. Together with nucleons, they can compose an atomic nuclei.

With the help of the simple method presented above, we have checked that the potentials of $M^0M^0$ and $\bar{M}^0\bar{M}^0$ interactions are attractive for the case of scalar meson exchange. It is repulsive for the case of vector meson exchange. The potentials of $M^0\bar{M}^0$ interactions have attractive asymptotes both for scalar and vector meson exchanges. Thus, the sign of potential for the cases of $M^0M^0$ and $\bar{M}^0\bar{M}^0$ scattering is determined by contributions of scalar and vector mesons. In the case of $M^0\bar{M}^0$ scattering, the total potential is attractive in all channels. This property leads to the effect of enhancement of annihilation cross section (Sommerfeld--Gamov--Sakharov effect~\cite{Som,Gam,Sak,Sakhenhance,ArbKo,SEeff}).

\subsection{The Properties of New Heavy Particles and Hadronic Dark Matter}

In this subsection, we consider the main properties of new hadrons $M^0$, $M^-$ and analyze the possibility that $M^0$ is stable and can be considered as DM candidate. We evaluate the mass
of the new quark $M_Q$ and mass splitting of the $M^-$ and $M^0$ mesons, $\Delta m = m^- - m^0$. Then, we take into account the standard electromagnetic and strong interactions of new hadrons which were described in a previous subsection. The properties of mesons $M=(M^0,M^-)$ are analogous to ones of standard heavy-light mesons. Let us consider the data on mass splitting in pairs $K=(K^0,K^{\pm})$, $D=(D^0,D^{\pm})$, and
$B=(B^0,B^{\pm})$. For the case of the mesons $K$ and $B$, which contain heavy down-type quarks, the~mass-splitting $\Delta m <0$. For the case of the up-type meson $D$, which contains heavy charm quark, $\Delta m>0$. The value $\delta m$ for all cases is $O(\mbox{MeV})$ and less. We take into consideration these data and assume that for the case of up-type new mesons
\begin{equation}\label{5.1}
\Delta m =m(M^-)-m(M^0)>0,\,\,\,\mbox{and}\,\,\, \Delta m=O(\text{MeV}).
\end{equation}

This assumption means that the neutral state $M^0=(\bar{U}u)$ is stable and can be considered as the DM candidate. The charged partner $M^-=(\bar{U}d)$ is unstable, if $\delta m>m_e$,
and has only one decay channel with small phase space in the final state:
\begin{equation}\label{5.2}
M^-\to M^0 (W^{-})^{*}\to M^0e^-\bar{\nu}_e,
\end{equation}
where $(W^{-})^{*}$ is a $W$-boson in intermediate state and $e^-$ is an electron. In the semileptonic decay (\ref{5.2}), stable antiquark $\bar{U}$ is considered as a spectator. The width of this decay is calculated in the form-factor approach. The expression for differential width is (see review by R. Kowalski in ref.~\cite{PDG})
\begin{equation}\label{5.3}
\dfrac{d\Gamma(m,\Delta m)}{d\kappa}=\dfrac{G^2_F}{48\pi^3}|U_{ud}|^2(m_-+m_0)^2m_0^3(\kappa^2-1)^{3/2}G^2(\kappa),
\end{equation}
where $m_-\approx m_0$, $\kappa =k^0/m_0\approx 1$ and $G(\kappa)\approx 1$ (HQS 
approximation,~\cite{PDG}). Note, the value $G(\omega)$ is equivalent to the normalized form-factor
$f_+(q)$. This form-factor in the vector dominance approach is usually defined by the pole expression $f_+(q)=f_+(0)/(1-q^2/m^2_v)$. So, the HQS approximation corresponds to the conditions $q^2\ll m^2_v$ and $f_+(0)\approx 1$ when $\kappa =k^0/m_0\approx 1$. The expression for the total width follows from Equation~(\ref{5.3}):
\begin{equation}\label{5.4}
\Gamma(m,\Delta m)\approx \dfrac{G^2_F|U_{ud}|^2 m^5_0}{12\pi^3}\int_1^{\kappa_m}(\kappa^2-1)^{3/2} d\kappa,
\end{equation}
where $\kappa_m=(m^2_0+m^2_-)/2m_0m_-$. After the integration, the expression (\ref{5.4}) can be written in a simple~form:
\begin{equation}\label{5.5}
\Gamma(\Delta m)\approx \dfrac{G_F^2}{60\pi^3}(\Delta m)^5.
\end{equation}

So, the width crucially depends on the mass splitting, $\Gamma\sim (\Delta m)^5$. It does not depend on the mass of heavy meson. In the interval $\Delta m=(1-10)\,\mbox{MeV}$ we get
following estimations:
\begin{equation}\label{5.6}
\Gamma\sim (10^{-29}-10^{-24})\,\mbox{GeV};\,\,\,\tau\sim (10^5-10^0)\,\mbox{s}.
\end{equation}

Thus, charged particle $M^-$ can be detected in the processes of $M^0 N$-collisions. This possibility was analyzed in ref.~\cite{10} (and references therein), where indirect experimental evidences for the presence of heavy charged metastable particles in cosmic rays were considered. Here, we should note that the scenario with a long-lived co-annihilation partner is considered in refs.~\cite{9a,Khoze}.

Further, we estimate the mass of new heavy hadrons in the scenario where they are interpreted as dark matter candidates. The data on the DM relic concentration lead to the equality
\begin{equation}\label{5.7}
(\sigma(M) v_r)^{exp}\approx 10^{-10}\,\,\text{GeV}^{-2}.
\end{equation}

In (\ref{5.7}), $M$ is the mass of new hadron.
From this equality we estimate the mass of the meson $M^0$. Note, the calculations
are done for the case of hadron-symmetric DM. To escape the problem with anomalous helium in this case, we expect $M_Q > 10$ TeV. Evaluation of the cross section $\sigma(M^0\bar{M}^0)$ is fulfilled in the approach
 $\sigma(M^0\bar{M}^0)\sim\sigma(U\bar{U})$, where $U$ is new heavy quark and we consider the light $u$-quark as a spectator. So, we estimate the cross section at the level of sub-processes
 with participation of a heavy quark, where the main contributions follow from sub-processes
 $U\bar{U}\to gg$ and $U\bar{U}\to q\bar{q}$. The expressions for these cross sections were represented in the third subsection
 (Equations~(\ref{2.11}) and~(\ref{2.12})) and we use their sum for evaluation of the total cross section. Thus, we estimate the mass $m(M^0)\approx M_U$ from the approximate equation
\begin{equation}\label{5.8}
(\sigma(M) v_r)^{exp}\approx \dfrac{44\pi}{9}\dfrac{\alpha^2_s}{M^2_U}.
\end{equation}

From (\ref{5.7}) and (\ref{5.8}) it follows that $m(M^0)=M\approx M_U\approx 20\,\mbox{TeV}$ at $\alpha_s=\alpha_s(M)$. This values are in accordance with the results in Ref.~\cite{Asadi} for the case of heavy WIMPonium.

Attractive potential of $M^0\bar{M}^0$-interaction, as was noted in the previous subsection, can increase annihilation cross section due to the light meson exchange at long distance.
This effect leads to the so-called Sommerfeld--Gamov--Sakharov (SGS) enhancement~\cite{SEeff}:
\begin{equation}\label{5.9}
\sigma(M) v_r=\sigma_0(M) v_r K(2\alpha/v_r).
\end{equation}

Here, $\sigma_0(M)$ is the initial cross section, $\alpha=g^2/4\pi$ is strong coupling which is defined in (\ref{4.2}). At~$m\ll M\approx M_U$, where $m$ is mass of intermediate mesons (the light force mediators), the SGS factor $K$ can be represented in the
form~\cite{SEeff}:
\begin{equation}\label{5.10}
K(2\alpha/v_r)=\dfrac{2\pi \alpha/v_r}{1-\exp(-2\pi \alpha/v_r)}\,.
\end{equation}

The light force carriers, in the case under consideration, are $\omega$- and $\rho$-mesons and $\alpha\sim 1$, so, from Equation~(\ref{5.10}) we get  the estimations $10^2 \lesssim K(2\alpha/v_r)/\pi\lesssim 10^3$ in the interval $10^{-2}>v_r>10^{-3}$. Thus, from the Equations~(\ref{5.8})--(\ref{5.10}) it follows that at $v_r\sim 10^{-2}$ the mass of new quark $M_U\sim 10^2\,\mbox{TeV}$. This value agrees with the estimations of the baryonic DM mass in~\cite{RanHuo} ($M\sim 100$ TeV). So, the value $M_U$ falls out from the mass range of the searches for anomalous hydrogen ($M_{max}\lesssim 1\,\mbox{TeV}$) and anomalous helium ($M_{max}\lesssim 10\,\mbox{TeV}$). In our estimations we take into account the light mesons only, ($m\ll M$). At~a~short distance, near $r\sim M^{-1}$, the exchange by heavy mesons is possible. The expression (\ref{5.10}), in this case, is not valid because $M_{\chi}\sim M_U$, where $M_{\chi}$ is the mass of heavy force mediators. For evaluation of SGS factor $K$ in this case, we use the numerical calculations in ref.~\cite{Cirelli}. From this work it follows that $K\approx 10$ in the interval $10^{-1}>v_r>10^{-3}$, and we get from (\ref{5.8}) and (\ref{5.9}) the estimation $M\approx 60\,\text{TeV}$ which does not crucially change the situation. Here, we should note that correct description of SGS requires taking account of bosons $Z$ and $W$ also. Thus, SGS effects are formed at various energies which correspond to various distances. So, this effect has a very complicated and vague nature (see also ref.~\cite{Blum}).

%===========================================================================
%===========================================================================
%===========================================================================

\section{Hypercolor Extensions of Standard Model}\label{hyper}

In this section, we consider some particular variants of models that extend SM by introducing an additional strong sector with heavy vector-like fermions, hyperquarks (H-quarks), charged under an H-color gauge group~\cite{Sundrum,Pasechnik:2013bxa,Pasechnik:2013kya,Lebiedowicz:2013fta,Pasechnik:2014ida,doi:10.1142/S0217751X17500427,2015JHEP...12..031A,2017PhRvD..95c5019A,2018JHEP...08..017B,2015JHEP...01..157A,2017JHEP...10..210M,Appelquist_ref}. Depending on H-quark quantum numbers, such models can encompass scenarios with composite Higgs doublets (see e.g.,~\cite{Cai:2018tet}) or a small mixing between fundamental Higgs fields of SM and composite hadron-like states of the new strong sector making the Higgs boson partially composite. Models of this class leave room for the existence of DM candidates whose decays are forbidden by accidental symmetries. Besides, H-color models comply well with electroweak  precision constraints, since H-quarks are assumed to be vector-like.

In the rest of this section, we briefly review one of the simplest realizations of the scenario described---models with two or three vector-like H-flavors confined by strong H-color force $\text{Sp}(2\chi_{\tilde{c}})$, $\chi_{\tilde{c}} \geqslant 1$. The models with H-color group SU(2)~\cite{2017PhRvD..95c5019A,Beylin:2016kga} are included as particular cases in this consideration due to isomorphism $\text{SU}(2) = \text{Sp}(2)$~\cite{2017PhRvD..95c5019A,Beylin:2016kga}. The global symmetry group of the strong sector with symplectic H-color group is larger than for the special unitary case---it is the group SU($2n_F$) broken spontaneously to Sp($2n_F$), with $n_F$ being a number of H-flavors. We posit that the extensions of SM under consideration preserve the elementary Higgs doublet in the set of Lagrangian field operators. This doublet mixes with H-hadrons, which makes the physical Higgs partially composite. Note also that the same coset SU($2n_F$)/Sp($2n_F$) can be used to construct composite two Higgs doublet model~\cite{Cai:2018tet} or little Higgs models~\cite{Low:2002ws,Csaki:2003si,Gregoire:2003kr,Han:2005dz,Brown:2010ke,Gopalakrishna:2015dkt}.

%===========================================================================
%===========================================================================
%===========================================================================

\subsection{\label{sec:lag/sym}Lagrangian and Global Symmetry of Symplectic QCD with $n_F =2,$ 3 Hyperquark Flavors}

In this section, we consider the simplest possibilities to extend the symmetry of SM, $G_\text{SM}$, by adding a symplectic hypercolor group, i.e., the gauge group of the extension under consideration is $G=G_\text{SM} \times \text{Sp}(2\chi_{\tilde{c}})$, $\chi_{\tilde{c}} \geqslant 1$. The model is postulated to have new degrees of freedom, six hyperquarks---Weyl fermions charged under H-color group. These fermions are assumed to form two weak doublets $Q_{\text{L}(A)}^{k\underline{k}}$ and two singlets $S_{\text{L}(A)}^{\underline{k}}$, $A=1,\,2$. In this paper, we underscore indices that are related to the H-color group $\text{Sp}(2\chi_{\tilde{c}})$; the normal Latin indices ($k$, $a$, etc.) are for the weak group $\text{SU}(2)_\text{L}$. The transformation law for the H-quarks is posited to be 
\begin{align}
(Q^{j\underline j}_{\text{L}(A)})'&{}=Q^{j\underline j}_{\text{L}(A)}-\dfrac{i}{2} g_1 Y_{Q(A)} \theta
Q^{j\underline j}_{\text{L}(A)}+\dfrac{i}{2}g_2 \theta_a
\tau_a^{jk}Q^{k\underline j}_{\text{L}(A)}+\dfrac{i}{2}g_{\tilde{c}}\theta_{\underline a}\lambda_{\underline a}^{\underline j \underline k}
Q^{j\underline k}_{\text{L}(A)}\, ,
\\
(S^{\underline j}_{\text{L}(A)})'&{}=S^{\underline j}_{\text{L}(A)}-ig_1 Y_{S(A)} \theta
S^{\underline j}_{\text{L}(A)}+\dfrac{i}{2}g_{\tilde{c}}\theta_{\underline a}\lambda_{\underline a}^{\underline j \underline k}S^{\underline k}_{\text{L}(A)}.
\end{align}

Here, $\theta$, $\theta_a$, $\theta_{\underline{a}}$ are transformation parameters of $\text{U}(1)_\text{Y}$, $\text{SU}(2)_\text{L}$, and $\text{Sp}(2\chi_{\tilde{c}})$ respectively; $\tau_a$ are the Pauli matrices; $\lambda_{\underline a}$, $\underline{a} = 1 \dots \chi_{\tilde{c}} (2\chi_{\tilde{c}} + 1)$ are $\text{Sp}(2\chi_{\tilde{c}})_{\tilde{c}}$ generators satisfying the relation 
\begin{align}\label{eq:scgr}
\lambda_{\underline a}^\text{T} \omega + \omega \lambda_{\underline a} = 0,
\end{align}
where $\text{T}$ stands for ``transpose'', $\omega$ is an antisymmetric $2\chi_{\tilde{c}} \times 2\chi_{\tilde{c}}$ matrix, $\omega^\text{T} \omega = 1$. From now on, $\text{SU}(2)_\text{L}$ and $\text{Sp}(2\chi_{\tilde{c}})_{\tilde{c}}$ indices are omitted if this does not lead to ambiguities. The relation \eqref{eq:scgr} and the analogous one holding true for the Pauli matrices of the weak group imply that the H-quarks are pseudoreal representations of the gauge symmetry groups of the model. This allows us to write the right-handed fields exhibiting transformation properties that are similar to those of the original left-handed ones:  
\begin{align}
Q_{\text{R}(A)}'&{} = \varepsilon \omega Q_{\text{L}(A)}{}^\text{C}
=Q_{\text{R}(A)} + \dfrac{i}{2} g_1 Y_{Q(A)} \theta
Q_{\text{R}(A)}+\dfrac{i}{2}g_2 \theta_a
\tau_a Q_{\text{R}(A)}+\dfrac{i}{2}g_{\tilde{c}}\theta_{\underline a}\lambda_{\underline a}
Q_{\text{R}(A)},
\\
S_{\text{R}(A)}'&{} = \omega S_{\text{L}(A)}{}^\text{C}
=S_{\text{R}(A)} + ig_1 Y_{S(A)} \theta
S_{\text{R}(A)}+\dfrac{i}{2}g_{\tilde{c}}\theta_{\underline a}\lambda_{\underline a}S_{\text{R}(A)},
\end{align}
where $\varepsilon=i\tau_2$.

The quantum numbers of the right-handed spinors $Q_{\text{R}(A)}$ and $S_{\text{R}(A)}$ are the same as the ones of the left-handed H-quarks except for the opposite-sign hypercharges. Therefore, setting $Y_{Q(1)}=-Y_{Q(2)}=Y_Q$ and  $Y_{S(1)}=-Y_{S(2)}=Y_S$, we obtain a doublet and a singlet of Dirac fields: 
\begin{align}
Q=Q_{\text{L}(1)} + Q_{\text{R}(2)},
\qquad
S=S_{\text{L}(1)} + S_{\text{R}(2)}.
\end{align}

These relations among hypercharges are also enforced independently by requiring cancellation of gauge anomalies.  

Finally, the Lagrangian of the SM extension invariant under $G=G_\text{SM} \times \text{Sp}(2\chi_{\tilde{c}})$ reads   
\begin{gather}\label{eq:LQS1}
    \mathscr{L} = \mathscr{L}_\text{SM} - \dfrac14 H^{\mu\nu}_{\underline{a}} H_{\mu\nu}^{\underline{a}} + i \bar Q \slashed{D} Q - m_Q \bar{Q} Q + i \bar S \slashed{D} S - m_S \bar{S} S + \delta\mathscr{L}_{\text{Y}},
\\
    D^\mu Q = \left[ \partial^\mu + \dfrac{i}{2} g_1 Y_Q B^\mu - \dfrac{i}{2} g_2 W_a^\mu \tau_a - \dfrac{i}2 g_{\tilde{c}} H^\mu_{\underline{a}} \lambda_{\underline{a}} \right] Q,
\\
    D^\mu S = \left[ \partial^\mu + i g_1 Y_S B^\mu - \dfrac{i}2 g_{\tilde{c}} H^\mu_{\underline{a}} \lambda_{\underline{a}} \right] S,
\end{gather}
where $H^\mu_{\underline{a}}$, $\underline{a} = 1 \dots \chi_{\tilde{c}} (2\chi_{\tilde{c}} + 1)$ are hypergluon fields and $H^{\mu\nu}_{\underline{a}}$ are their strength tensors. Contact Yukawa couplings $\delta\mathscr{L}_{\text{Y}}$ of the H-quarks and the SM Higgs doublet $\mathscr{H}$ are permitted in the model if the hypercharges satisfy an additional linear relation: 
\begin{gather}\label{eq:LQHS}
    \delta\mathscr{L}_{\text{Y}} = y_\text{L} \left( \bar{Q}_\text{L} \mathscr{H} \right) S_\text{R} + y_\text{R} \left( \bar{Q}_\text{R} \varepsilon \bar{\mathscr{H}} \right) S_\text{L} + \text{h.c.} \quad \text{ for } \dfrac{Y_Q}{2}-Y_S = +\dfrac12;
\\
    \delta\mathscr{L}_{\text{Y}} = y_\text{L} \left( \bar{Q}_\text{L} \varepsilon \bar{\mathscr{H}} \right) S_\text{R} + y_\text{R} \left( \bar{Q}_\text{R} \mathscr{H} \right) S_\text{L} + \text{h.c.} \quad \text{ for } \dfrac{Y_Q}{2}-Y_S = -\dfrac12.
\end{gather}

The model can be reconciled with the electroweak precision constraints quite easily, since H-quarks are vector-like, i.e., their electroweak interactions are chirally symmetric in this scheme. Besides, this allows us to introduce explicit gauge-invariant Dirac mass terms for H-quarks.  

It is easy to prove that the kinetic terms of H-quarks $Q$ and $S$ in the Lagrangian \eqref{eq:LQS1} can be rewritten in terms of a left-handed sextet as follows: 
\begin{gather}\label{eq:LP}
    \delta \mathscr{L}_\text{H-quarks, kin} = i \bar P_\text{L} \slashed{D} P_\text{L} ,
    \qquad P_\text{L} = \left(  Q_{\text{L}(1)}^\text{T}, \; Q_{\text{L}(2)}^\text{T}, \; S_{\text{L}(1)}, \; S_{\text{L}(2)} \right)^\text{T},
%    \qquad P_\text{L} = \begin{pmatrix} Q_{\text{L}(1)} \\ Q_{\text{L}(2)} \\ S_{\text{L}(1)} \\ S_{\text{L}(2)} \end{pmatrix},
    \\\label{eq:covdcur}
    D^\mu P_\text{L} = \left[ \partial^\mu + i g_1 B^\mu \left( Y_Q \Sigma_Q + Y_S \Sigma_S \right) - \dfrac{i}{2} g_2 W_a^\mu \Sigma^a_W - \dfrac{i}2 g_{\tilde{c}} H^\mu_{\underline{a}} \lambda_{\underline{a}} \right] P_\text{L},
\\ \label{eq:SigmaQSW}
    \Sigma_Q = \dfrac12 \begin{pmatrix} 1 & 0 & 0 \\ 0 & -1 & 0 \\ 0 & 0 & 0 \end{pmatrix},
\qquad
    \Sigma_S = \begin{pmatrix} 0 & 0 & 0 \\ 0 & 0 & 0 \\ 0 & 0 & \tau_3 \end{pmatrix},
\qquad
    \Sigma^a_W = \begin{pmatrix} \tau_a & 0 & 0 \\ 0 & \tau_a & 0 \\ 0 & 0 & 0 \end{pmatrix}.
\end{gather}

In the limit of vanishing electroweak interactions, $g_1=g_2=0$, this Lagrangian is invariant under a global SU(6) symmetry, which is dubbed as the Pauli--G\"{u}rsey  symmetry sometimes~\cite{Pauli1957Conservation, Gursey1958Relation}:
\begin{gather}\label{eq:PLTP}
    P_\text{L} \to U P_\text{L}, \qquad U \in \text{SU}(6).
\end{gather}

 The subgroups of the SU(6) symmetry include:  
\begin{itemize}
\item the chiral symmetry $\text{SU}(3)_\text{L} \times \text{SU}(3)_\text{R}$,  
\item SU(4) subgroup corresponding to the two-flavor model without singlet H-quark $S$,  
\item two-flavor chiral group $\text{SU}(2)_\text{L} \times \text{SU}(2)_\text{R}$, which is a subgroup of both former subgroups.  
\end{itemize}

The global symmetry is broken both explicitly and dynamically:  
\begin{itemize}
\item explicitly---by the electroweak and Yukawa interactions, \eqref{eq:covdcur} and \eqref{eq:LQHS}, and the H-quark masses;  
\item dynamically---by H-quark condensate  ~\cite{Vysotskii1985Spontaneous,Verbaarschot2004Supersymmetric}:
\begin{gather}\label{eq:LTQ}
    \langle \bar QQ + \bar SS \rangle = \dfrac12 \langle  \bar P_\text{L} M_0 P_\text{R} + \bar P_\text{R} M_0^\dagger P_\text{L} \rangle,
\qquad P_\text{R}  = \omega P_\text{L}{}^\text{C},
    \qquad  M_0 = \begin{pmatrix} 0 & \varepsilon & 0\\ \varepsilon & 0 & 0 \\ 0 & 0 & \varepsilon \end{pmatrix}.
\end{gather}
\end{itemize}

The condensate \eqref{eq:LTQ} is invariant under $\text{Sp}(6) \subset \text{SU}(6)$ transformations $U$ that satisfy a condition
\begin{gather}\label{eq:symprel}
    U^\text{T} M_0 + M_0 U =0,
\end{gather}
i.e., the global SU(6) symmetry is broken dynamically to its Sp(6) subgroup. The mass terms of H-quarks in \eqref{eq:LQS1} could break the symmetry further to  $\text{Sp(4)}\times\text{Sp}(2)$:
\begin{gather}\label{eq:massterm}
    \delta \mathscr{L}_\text{H-quarks, masses} = -\dfrac12 \bar P_\text{L} M_0' P_\text{R} + \text{h.c.},
\qquad M'_0 = -M'_0{}^\text{T} = \begin{pmatrix} 0 & m_Q \varepsilon & 0\\ m_Q \varepsilon & 0 & 0 \\ 0 & 0 & m_S \varepsilon \end{pmatrix}.
\end{gather}

The case of a two-flavor model (without the singlet H-quark) is completely analogous to the three-flavor model but is simpler than latter one---the global SU(4) symmetry is broken dynamically to its Sp(4) subgroup by the condensate of doublet H-quarks; the Lagrangian of the model is obtained from the one given by Equations~\eqref{eq:LQS1}--\eqref{eq:SigmaQSW} by simply setting to zero all terms with the H-quark $S$.

%=============================================================================
%=============================================================================
%=============================================================================

\subsection{\label{sec:LSM}Linear Sigma Model As an Effective Field Theory of Constituent H-Quarks}

Now, we proceed to construct a linear $\sigma$-model for interactions of constituent H-quarks. The~Lagrangian of the model consists of kinetic terms for the constituent fermions and the lightest (pseudo)scalar composite states, Yukawa terms for the interactions of the (pseudo)scalars with the fermions, and a potential of (pseudo)scalar self-interactions $U_\text{scalars}$. The Lagrangian reads
\begin{gather}\label{eq:ctq}
    \mathscr{L}_{\text{L}\sigma} = \mathscr{L}_\text{H-quarks} + \mathscr{L}_\text{Y} + \mathscr{L}_\text{scalars},
\\[3mm]
    \mathscr{L}_\text{H-quarks}  = i \bar P_\text{L} \slashed{D} P_\text{L},
\qquad
    \mathscr{L}_\text{Y} =  -\sqrt2 \varkappa \left( \bar P_\text{L} M P_\text{R} + \bar P_\text{R} M^\dagger P_\text{L} \right),
\\ \label{eq:L-scalars}
    \mathscr{L}_\text{scalars} =  D_\mu \mathscr{H}^\dagger \cdot D^\mu \mathscr{H} + \Tr D_\mu M^\dagger \cdot D^\mu M - U_\text{scalars},
\end{gather}

Here, $\varkappa$ is a coupling constant; $M$ is a complex antisymmetric $2n_F \times 2n_F$ matrix of (pseudo)scalar fields; the multiplets $P_\text{L, R}$ correspond now to the constituent H-quarks but retain all the definitions and properties of the fundamental multiplets described in the previous section. The fields transform under the global symmetry SU($2n_F$) as follows:
\begin{gather}\label{eq:M_transf}
    M \to UMU^\text{T}, \qquad P_\text{L} \to U P_\text{L}, \qquad P_\text{R} \to \bar{U} P_\text{R}, \qquad U \in \text{SU}(2n_F),
\end{gather}
where $\bar{U}$ designates the complex conjugate of the matrix $U$. Note also that the model comprises of  the fundamental (not composite) Higgs doublet $\mathscr{H}$ of SM.

It is postulated that the interactions of the constituent H-quarks with the gauge bosons are the same as for the fundamental H-quarks. This and the transformation laws \eqref{eq:M_transf} define the covariant derivative for the scalar field $M$. The complete set of covariant derivatives present in the Lagrangian~\eqref{eq:ctq} is as~follows:
\begin{align}
	 D_\mu \mathscr{H} = \left[ \partial_\mu + \dfrac{i}{2} g_1 B_\mu - \dfrac{i}{2} g_2 W_\mu^a \right] \mathscr{H},
\qquad
	D^\mu P_\text{L} = \left[ \partial^\mu + i g_1 B^\mu \left( Y_Q \Sigma_Q + Y_S \Sigma_S \right) - \dfrac{i}{2} g_2 W_a^\mu \Sigma^a_W \right] P_\text{L},
 \label{eq:H-cd}
\end{align}
\begin{align}
    D_\mu M = \partial_\mu M
         + i Y_Q g_1 B_\mu (\Sigma_Q M + M \Sigma_Q^\text{T} )
         + i Y_S g_1 B_\mu (\Sigma_S M + M \Sigma_S^\text{T} )
    - \dfrac{i}{2} g_2 W_\mu^a (\Sigma^a_W M + M \Sigma^{a\text{T}}_W{} ),
\label{eq:M-cd}
\end{align}
where the matrices $\Sigma_Q$, $\Sigma_S$, $\Sigma^a_W$, $a=1,\,2,\,3$ are defined by Equation~\eqref{eq:SigmaQSW}.

%=====================================================================================
%=====================================================================================
%=====================================================================================

\subsubsection{Interactions of the Constituent H-Quarks with H-Hadrons and the Electroweak Gauge Bosons} 

In the case of $n_F=3$, the field $M$ can be expanded in a basis of fourteen ``broken'' generators $\beta_a$ of the global symmetry group SU(6):
\begin{align}
    M &{}= \left[ \dfrac1{2\sqrt{n_F}} (A_0+iB_0) I + (A_a+iB_a) \beta_a \right] M_0
\notag\\[2ex]
    &{}= \dfrac12 \begin{pmatrix} \bar{A} \varepsilon & \left[ \dfrac{1}{\sqrt{n_F}} \sigma + \dfrac{1}{\sqrt{2n_F}} f + \dfrac{1}{\sqrt2} a_a \tau_a \right] \varepsilon & K^\star \varepsilon \\
                                         \left[ \dfrac{1}{\sqrt{n_F}} \sigma + \dfrac{1}{\sqrt{2n_F}} f - \dfrac{1}{\sqrt2} a_a \tau_a \right] \varepsilon & A \varepsilon & \varepsilon \bar{K}^\star \\
                                         K^{\star\dagger} \varepsilon & \varepsilon K^{\star\text{T}} & \dfrac{1}{\sqrt{n_F}} \left( \sigma - \sqrt2 f \right) \varepsilon
                \end{pmatrix}
\notag\\ \label{eq:Mexp}
        &{}\hphantom{=}+ \dfrac{i}2 \begin{pmatrix} \bar{B} \varepsilon & \left[ \dfrac{1}{\sqrt{n_F}}\eta + \dfrac{1}{\sqrt{2n_F}} \eta' + \dfrac{1}{\sqrt2} \pi_a \tau_a \right] \varepsilon & K \varepsilon \\
                                         \left[ \dfrac{1}{\sqrt{n_F}} \eta + \dfrac{1}{\sqrt{2n_F}} \eta' - \dfrac{1}{\sqrt2} \pi_a \tau_a \right] \varepsilon & B \varepsilon & \varepsilon \bar{K} \\
                                         K^\dagger \varepsilon & \varepsilon K^\text{T} & \dfrac{1}{\sqrt{n_F}} \left( \eta - \sqrt2 \eta' \right) \varepsilon
                \end{pmatrix} ,
\end{align}

Here, $I$ is the identity matrix and new scalar fields are defined as follows:
\begin{gather}
    \sigma = A_0, \quad \eta = B_0, \quad
    f = A_6, \quad \eta' = B_6, \quad
    a_a = A_{a+2}, \quad \pi_a = B_{a+2}, \quad
    a=1,\,2,\,3,
    \notag\\
    A = \dfrac1{\sqrt2} ( A_1 + i A_2 ), \quad
    B = \dfrac1{\sqrt2} ( B_1 + i B_2 ),
    \notag\\
    K^\star = \dfrac12 \left[ A_{10} + i A_{14} + ( A_{6+a} + i A_{10+a} ) \tau_a \right],
    \qquad
    K = \dfrac12 \left[ B_{10} + i B_{14} + ( B_{6+a} + i B_{10+a} ) \tau_a \right].
\notag\end{gather}

The generators $\beta_a$ are defined in the Appendix \ref{app:gens}. A bar over a scalar field denote the complex conjugate of the field operator. In the case of $n_F=2$, we should substitute the identity matrix $I$ in Equation~\eqref{eq:Mexp} by the diagonal matrix $\mathop{\text{diag}}(1,1,1,1,0,0)$ and take into account just the first five of generators $\beta_a$, i.e.,\ $A_a=0=B_a$ for $a=6, \dots 14$ or, equivalently, $K=0=K^\star$ and $f=0=\eta'$. In other words, only the upper left $4 \times 4$ block of the matrix \eqref{eq:Mexp} remains under consideration, while all other its elements are set to zero.

Assuming that the singlet meson $\sigma$ develops a v.e.v. $u$, $\sigma = u + \sigma'$, and inserting the representation~\eqref{eq:Mexp} into the Lagrangian \eqref{eq:ctq} of the sigma model,  we arrive at the following form of the Lagrangian:
\begin{gather}
%\begin{gather}%\label{eq:}
    \mathscr{L}_\text{H-quarks} + \mathscr{L}_\text{Y} = i \bar Q \slashed{D} Q + i \bar S \slashed{D} S - \varkappa u \left( \bar Q Q + \bar S S \right)
    \notag\\[2ex]
    -\varkappa \bar{Q} \left[ \sigma' + \dfrac{1}{\sqrt3} f + i \left( \eta + \dfrac{1}{\sqrt3} \eta' \right) \gamma_5
     + \left( a_a + i \pi_a \gamma_5 \right) \tau_a
    \right]  Q
%   \notag\\
    -\varkappa \bar{S} \left[ \sigma' - \dfrac{2}{\sqrt3} f
 + i \left( \eta - \dfrac{2}{\sqrt3} \eta' \right) \gamma_5 \right] S
   \notag\\[2ex]
    -\sqrt2 \varkappa \left[ \left( \bar Q \mathscr{K}^\star \right) S + i \left( \bar Q \mathscr{K} \right) \gamma_5 S + \text{h.c.} \right]
	 -\sqrt2 \varkappa \left[ \left( \bar Q \mathscr{A} \right) \omega S^\text{C}
    + i \left( \bar Q \mathscr{B} \right) \gamma_5 \omega S^\text{C} + \text{h.c.} \right]
   \notag\\ [1ex]
    -\dfrac{\varkappa}{\sqrt2}  \left( A \bar Q \varepsilon \omega Q^\text{C}
    +iB \bar Q \gamma_5 \varepsilon \omega Q^\text{C} + \text{h.c.} \right) ,
\label{eq:LSM-lagr-QS-phys}
\end{gather}
\begin{equation}
    D_\mu Q = \partial_\mu Q + \dfrac{i}2 g_1 Y_Q B_\mu Q - \dfrac{i}2 g_2 W_\mu^a \tau_a Q,
    \qquad
    D_\mu S = \partial_\mu S + i g_1 Y_S B_\mu S,\label{eq:QS-cd}
\end{equation}
where $\mathscr{K}^\star$, $\mathscr{K}$ and $\mathscr{A}$, $\mathscr{B}$ are $\text{SU}(2)_\text{L}$ doublets of H-mesons and H-diquarks (H-baryons) respectively:
\begin{gather}%\label{eq:}
    \mathscr{K}^\star = \dfrac{1}{\sqrt2} \left( R_1 + i R_2 \right),
\,
    \mathscr{K} = \dfrac{1}{\sqrt2} \left( S_1 + i S_2 \right),
\,
    \mathscr{A} = \dfrac{1}{\sqrt2} \varepsilon \left( \bar R_1 + i \bar R_2 \right),
\,
    \mathscr{B} = \dfrac{1}{\sqrt2} \varepsilon \left( \bar S_1 + i \bar S_2 \right),
\\
    R_1 = \dfrac{1}{\sqrt2} \begin{pmatrix} A_{10} + i A_{13} \\ -A_{12} + i A_{11} \end{pmatrix},
\qquad
    R_2 = \dfrac{1}{\sqrt2} \begin{pmatrix} A_{14} - i A_{9} \\ A_{8} - i A_{7} \end{pmatrix},
\\
    S_1 = \dfrac{1}{\sqrt2} \begin{pmatrix} B_{10} + i B_{13} \\ -B_{12} + i B_{11} \end{pmatrix},
\qquad
    S_2 = \dfrac{1}{\sqrt2} \begin{pmatrix} B_{14} - i B_{9} \\ B_{8} - i B_{7} \end{pmatrix}.
\end{gather}

The Lagrangian for the case of a two-flavor model, $n_F = 2$, is obtained by simply neglecting all terms with the singlet H-quark $S$ in Equation~\eqref{eq:LSM-lagr-QS-phys}.

%============================================================================================
%============================================================================================
%============================================================================================

\subsubsection{Interactions of the (Pseudo)scalar Fields with the Electroweak Gauge Bosons}

The kinetic terms of the lightest H-hadrons in the Lagrangian \eqref{eq:L-scalars} can be put into the following~form:
\begin{align}\label{eq:TM}
    \mathscr{T}_\text{scalars}  = \dfrac12 \sum_\varphi D_\mu \varphi \cdot D^\mu \varphi  + \sum_\Phi \left( D_\mu \Phi \right)^\dagger D^\mu \Phi
	 + D_\mu \bar A \cdot D^\mu A + D_\mu \bar B \cdot D^\mu B ,
\end{align}
where $\varphi = h$, $h_a$, $\pi_a$, $a_a$, $\sigma$, $f$, $\eta$, $\eta'$ are singlet and triplet fields, $\Phi = \mathscr{K}$, $\mathscr{K}^\star$, $\mathscr{A}$, $\mathscr{B}$ are doublets. The~fields $h$ and $h_a$, $a=1$, 2, 3 are components of the fundamental Higgs doublet
\begin{gather}
   \mathscr{H} = \dfrac{1}{\sqrt2} \begin{pmatrix} h_2+ih_1 \\ h-ih_3\end{pmatrix}.
\end{gather}

All covariant derivatives in the Lagrangian \eqref{eq:TM} follow directly from the covariant derivatives of the fields $\mathscr{H}$ \eqref{eq:H-cd} and $M$ \eqref{eq:M-cd}:
\begin{gather}\label{eq:dpi}
	D_\mu h
	= \partial_\mu h
             +\dfrac12 (g_1 \delta_3^a B_\mu +g_2 W_\mu^a ) h_a,
	\qquad
	D_\mu \phi
	= \partial_\mu \phi, \quad \phi=\sigma, \, f, \, \eta, \, \eta',
	\\[1ex]
    D_\mu h_a
        =\partial_\mu h_a
             -\dfrac12 (g_1 \delta_3^a B_\mu +g_2 W_\mu^a ) h
             -\dfrac12 e_{abc} (g_1 \delta_3^b B_\mu -g_2 W_\mu^b ) h_c,
	\\[1ex]
    D_\mu \pi_a = \partial_\mu \pi_a + g_2 e_{abc} W_\mu^b \pi_c ,
    \qquad
    D_\mu a_a = \partial_\mu a_a + g_2 e_{abc} W_\mu^b a_c,
    \\[1ex]
    D_\mu A = \partial_\mu A + i g_1 Y_Q B_\mu A,
    \qquad
    D_\mu B = \partial_\mu B + i g_1 Y_Q B_\mu B,
    \\[1ex]
    D_\mu \mathscr{K} = \left[ \partial_\mu + i g_1 \left( \dfrac{Y_Q}{2}-Y_S \right ) B_\mu - \dfrac{i}{2} g_2 W_\mu^a \tau^a \right] \mathscr{K},
	\\[1ex]
	D_\mu \mathscr{K}^\star = D_\mu \mathscr{K} \bigr|_{\mathscr{K} \to \mathscr{K}^\star},
	\quad
	D_\mu \mathscr{A} = D_\mu \mathscr{K} \biggr|{}_{ \begin{subarray}{l} \mathscr{K} \to \mathscr{A} \\ Y_S \to -Y_S \end{subarray} },
	\quad
	D_\mu \mathscr{B} = D_\mu \mathscr{K} \biggr|{}_{ \begin{subarray}{l} \mathscr{K} \to \mathscr{B} \\ Y_S \to -Y_S \end{subarray} }.
\end{gather}

\begin{table}
\caption{The lightest (pseudo)scalar H-hadrons in $\text{Sp}(2\chi_{\tilde{c}})$ model with two and three flavors of H-quarks (in the limit of vanishing mixings). The lower half of the table lists the states present only in the three-flavor version of the model. $T$ is the weak isospin. $\tilde G$ denotes hyper-$G$-parity of a state (see~Section~\ref{sec:vars}). $\tilde B$ is the H-baryon number. $Q_\text{em}$ is the electric charge (in units of the positron charge $e=|e|$). The H-quark charges are $Q^U_\text{em}  = (Y_Q+1)/2$, $Q^D_\text{em} = (Y_Q-1)/2$, and $Q^S_\text{em} = Y_S$, which is seen from \eqref{eq:QS-cd}.}
\centering\renewcommand{\arraystretch}{1.9}
\begin{ruledtabular}
{\begin{tabular}{ccccccccc}
\textbf{State} & $$ & \textbf{H-Quark Current} & $$ & $\bm{T^{\tilde G}(J^{PC})}$ &  & $\bm{\tilde{B}}$ &  & $\bm{Q_\text{\textbf{em}}}$ \\
\hline
$\sigma$ &  & $\bar Q Q + \bar SS$ &  & $0^+(0^{++})$ & $$ & 0 & $$ & 0 \\
$\eta$ &  & $i \left( \bar Q \gamma_5 Q + \bar S \gamma_5 S \right)$ &  & $0^+(0^{-+})$ & $$ & 0 & $$ & 0 \\
$ a_k$ &  & $\bar Q \tau_k Q$ &  & $1^-(0^{++})$ & $$ & 0 & $$ & $\pm 1$, 0 \\
$\pi_k$ &  & $i \bar Q \gamma_5 \tau_k Q$ &  & $1^-(0^{-+})$ & $$ & 0 & $$ & $\pm 1$, 0 \\
$A$ &  & $\bar Q^\text{C} \varepsilon \omega Q$ &  & $0^{\hphantom{+}}(0^{-\hphantom{+}})$ &  & 1 &  & $Y_Q$ \\
$B$ &  & $i \bar Q^\text{C}  \varepsilon \omega \gamma_5 Q$ &  & $0^{\hphantom{+}}(0^{+\hphantom{+}})$ & $$ & 1 & $$ & $Y_Q$ \\
\hline
$f$ &  & $\bar Q Q -2 \bar SS$ & & $0^+(0^{++})$ & $$ & 0 & $$ & 0 \\
$\eta'$ &  & $i \left( \bar Q \gamma_5 Q - 2 \bar S \gamma_5 S \right)$ &  & $0^+(0^{-+})$ & $$ & 0 & $$ & 0 \\
$\mathscr{K}^\star$ &  & $\bar S Q$ &  & $\dfrac12^{\hphantom{+}}(0^{+\hphantom{+}})$ & $$ & 0 & $$ & $Y_Q/2-Y_S \pm 1/2$ \\
$\mathscr{K}$ &  & $i \bar S \gamma_5 Q$ &  & $\dfrac12^{\hphantom{+}}(0^{-\hphantom{+}})$ & $$ & 0 & $$ & $Y_Q/2-Y_S \pm 1/2$ \\
$\mathscr{A}$ &  & $\bar S^\text{C} \omega Q$ &  & $\dfrac12^{\hphantom{+}}(0^{-\hphantom{+}})$ & $$ & 1 & $$ & $Y_Q/2+Y_S \pm 1/2$ \\
$\mathscr{B}$ &  & $i \bar S^\text{C} \omega \gamma_5 Q$ &  & $\dfrac12^{\hphantom{+}}(0^{+\hphantom{+}})$ & $$ & 1 & $$ & $Y_Q/2+Y_S \pm 1/2$
\end{tabular} \label{tab:H-hadrons} }
\end{ruledtabular}
\end{table}

%============================================================================================
%============================================================================================
%============================================================================================

\subsubsection{Self-Interactions and Masses of the (Pseudo)Scalar Fields}

The potential of spin-0 fields---the Higgs boson and (pseudo)scalar H-hadrons---can be written as~follows:
\begin{gather}\label{eq:U}
    U_\text{scalars} = \sum_{i=0}^4 \lambda_i I_i + \sum_{0 = i \leqslant k = 0}^3 \lambda_{ik} I_i I_k.
\end{gather}

Here, $I_i$, $i=0$, 1, 2, 3, 4 are the lowest dimension invariants
\begin{gather}\label{eq:Uinvs}
    I_0 = \mathscr{H}^\dagger \mathscr{H},
    \quad
    I_1 = \Tr \left( M^\dagger M \right),
    \quad
    I_2 = \mathop{\mathrm{Re}} \mathop{\mathrm{Pf}} M,
    \quad
    I_3 = \mathop{\mathrm{Im}} \mathop{\mathrm{Pf}} M,
    \quad
    I_4 = \Tr \left[ \left( M^\dagger M \right)^2 \right].
\end{gather}

The Pfaffian of $M$ is defined as
\begin{gather}\label{eq:Pfaffian}
    \mathop{\mathrm{Pf}} M = \dfrac1{2^2 2!} \varepsilon_{abcd} M_{ab} M_{cd} \quad \text{ for } n_F=2,
    \quad
    \mathop{\mathrm{Pf}} M = \dfrac1{2^3 3!} \varepsilon_{abcdef} M_{ab} M_{cd} M_{ef}  \quad \text{ for } n_F=3,
\end{gather}
where $\varepsilon$ is the $2n_F$-dimensional Levi--Civita symbol ($\varepsilon_{12\dots (2n_F)}=+1$). We consider only renormalizable part of the potential \eqref{eq:U} permitted by the symmetries of the model. This implies that $\lambda_{i2} = \lambda_{i3} = 0$ for all $i$ if $n_F = 3$. Besides, the invariant $I_3$ is CP odd, i.e., $\lambda_{3} = 0$ as well as $\lambda_{i3} = 0$ for $i=0$, 1, 2. In the two-flavor model, one of the terms in the potential \eqref{eq:U} is redundant because of the identity
\begin{gather}
    I_1^2 -4 I_2^2 -4 I_3^2 -2 I_4 =0
\end{gather}
that holds for $n_F = 2$. To take this into account, we set $\lambda_{22} = 0$. (As it is mentioned above, $\lambda_{22}$ is also set to zero for $n_F = 3$, since we consider only renormalizable interactions.)

In the case of vanishing Yukawa couplings $y_\text{L} = y_\text{R} = 0$, the tadpole equations for v.e.v.'s $v = \langle h \rangle \neq 0$ and $u = \langle \sigma \rangle \neq 0$  read
\begin{gather}\label{eq:tadpoles} 
    \mu_0^2 = \lambda_{00} v^2 + \dfrac12 \Lambda_{01} u^2,
    \quad
    \mu_1^2 = \dfrac12 \Lambda_{01} v^2 + \Lambda_{11} u^2  -\dfrac{\left(4-n_F \right) n_F \lambda_2 u^{n_F-2}}{2 \left(2\sqrt{n_F}\right)^{n_F}} + \dfrac32 \zeta \dfrac{\langle \bar Q Q + \bar S S \rangle}{u},
\end{gather}
where 
\begin{gather}
	\mu_0^2 = -\lambda_0,
	\qquad
	\mu_1^2 = -\lambda_1,
	\qquad
	\Lambda_{01} = \lambda_{01} - \dfrac14 \lambda_{02},
	\\
	\Lambda_{11} = \lambda_{11} + \dfrac{\lambda_4}{2n_F} - \dfrac14 \lambda_{12} - \dfrac{n_F (n_F-2) \lambda_2}{2 \left(2\sqrt{n_F}\right)^{n_F} u^{4-n_F}}  - \dfrac{\zeta \langle \bar Q Q + \bar S S \rangle}{2u^3} .
\end{gather}

The condition of vacuum stability requires that the following inequalities hold:
\begin{gather}\label{eq:vacineq} 
	\lambda_{00} > 0,
    \qquad
	\Lambda_{11} > 0,
    \qquad
    4 \lambda_{00} \Lambda_{11} - \Lambda_{01}^2 > 0.
\end{gather}

The effects of explicit breaking of the SU($2n_F$) global symmetry can be communicated to the effective fields by different non-invariant terms in the Lagrangian~\cite{1979PhRvC..19.1965C,2000PhRvC..61b5205D,1996NuPhA.603..239D}. Here, we use the most common one which is a tadpole-like term $\mathscr{L}_\text{SB} = -\zeta \langle \bar Q Q + \bar S S \rangle (u+\sigma')$, with the parameter $\zeta$ being proportional to the current mass $m_Q$ of the H-quarks  (see~\cite{1969RvMP...41..531G,2013PhRvD..87a4011P}, for example).

Tree masses of the (pseudo)scalars:
\begin{equation}
    m_{\sigma,H}^2 = \lambda_{00} v^2 + \Lambda_{11} u^2 \pm \sqrt{(\lambda_{00} v^2 - \Lambda_{11} u^2)^2 + \Lambda_{01}^2 v^2 u^2},
\end{equation}  
\begin{align}
    m_\pi^2 = m_{\eta'}^2 = m_B^2 = m_\mathscr{K}^2 = m_\mathscr{B}^2 = -\dfrac{\zeta \langle \bar Q Q + \bar S S \rangle}{u},\;
	m_\eta^2 = \left\{ \begin{alignedat}{2} & \dfrac12 \left( \dfrac14 \lambda_{33} - \lambda_4 \right) u^2 + m_a^2 \qquad & \text{for } n_F=2, \\ &\dfrac{\sqrt{3}}{24} \lambda_{2} u -\dfrac13 \lambda_4 u^2  +m_a^2 &\text{for } n_F=3, \end{alignedat} \right. 
 \end{align}  
\begin{equation}
    m_{f}^2 = m_\mathscr{K^\star}^2 = m_\mathscr{A}^2  = \dfrac{\sqrt{3}}{12} \lambda_{2} u + \dfrac13 \lambda_{4} u^2  -\dfrac{\zeta \langle \bar Q Q + \bar S S \rangle}{u},
 \end{equation}  
\begin{equation}
	m_a^2 = m_A^2 = \left\{ \begin{alignedat}{2} &  \dfrac12 \left( \lambda_4 + \dfrac12 \lambda_{12} \right) u^2 + \dfrac14 \lambda_{02} v^2 + \dfrac12 \lambda_2 -\dfrac{\zeta \langle \bar Q Q \rangle}{u} \qquad & \text{for } n_F=2, \\ &m_f^2 &\text{for } n_F=3. \end{alignedat} \right.
\end{equation}

For all $n_F$, the model involves a small mixing of the fundamental Higgs and H-meson $\sigma'$, which makes the Higgs partially composite:
\begin{gather}
    h = \cos\theta_s H - \sin\theta_s \sigma,
    \qquad
    \sigma' = \sin\theta_s H + \cos\theta_s \sigma,
    \\[3mm]
    \tan2\theta_s = \dfrac{\Lambda_{01} v u }{\lambda_{00} v^2 - \Lambda_{11} u^2},
    \qquad
    \sgn \sin\theta_s = -\sgn \Lambda_{01},
\end{gather}
where $H$ and $\sigma$ are physical fields.

%=============================================================================
%=============================================================================
%=============================================================================

\subsubsection{\label{sec:vars}Accidental Symmetries}

If the hypercharges of H-quarks are set to zero, the Lagrangian \eqref{eq:LQS1} is invariant under an additional symmetry---hyper $G$-parity~\cite{2010PhRvD..82k1701B,Antipin:2015xia}:
\begin{align}\label{eq:HGconjugation}
	Q^{\tilde{\text{G}}} = \varepsilon \omega Q^\text{C},
	\qquad
	S^{\tilde{\text{G}}} = \omega S^\text{C}.
\end{align}

Since H-gluons and all SM fields are left intact by \eqref{eq:HGconjugation}, the lightest $\tilde G$-odd H-hadron becomes stable. It~happens to be the neutral H-pion $\pi^0$.

Besides, the numbers of doublet and singlet quarks are conserved in the model \eqref{eq:LQS1}, because of the two global U(1) symmetry groups of the Lagrangian. This makes two H-baryon states stable---the neutral singlet H-baryon $B$ and the lightest state in doublet $\mathscr{B}$, which carries a charge of $\pm 1/2$.

%=============================================================================
%=============================================================================
%=============================================================================

\subsection{Physics and Cosmology of Hypercolor SU(4) and SU(6) Models}

So, in the simplest case of zero hypercharge, it is possible to consider some experimentally observed consequences of SU(4) minimal model~\cite{Beylin:2016kga}. As it is seen from above, even in the minimal scenario of this type of hypercolor extension, there emerges a significant number of additional degrees of freedom. These new states, such as pNG or other hyperhadrons, can be detected in reactions at the collider at sufficiently high energies.

Readers should be reminded here of several papers that concern the formulation and construction of a vector-like hypercolor scheme~\cite{Sundrum,Pasechnik:2013bxa}. In addition to the awareness of the original ideology, which made it possible to avoid known difficulties of Technicolor, main potentially observable consequences of this type of the SM extension were analyzed qualitatively and quantitatively~\cite{Pasechnik:2013kya,Pasechnik:2014ida,doi:10.1142/S0217751X17500427,Appelquist_ref}. Also in these articles, the possibilities of the hypercolor models for explaining the nature of DM particles were discussed in detail. Namely, this extension of the SM offers several different options as DM candidates with specific features and predictions. Some of these scenarios will be discussed in more detail below.

The vector-like hypercolor model  contains two different scalar states with zero (or small) mixing, the Higgs boson and  $\tilde{\sigma}$, and possibility to analyze quantitatively an effect of this mixing tends to zero. Then, we should hope to find some New Physics signals not in channels with the Higgs boson, but from  production and decays of $\tilde{\sigma}$-meson (as shown by experimental data at the LHC, almost all predictions of the SM for the Higgs boson production cross sections in different channels as well as the widths of various modes of its decay are confirmed). Interestingly, the fluke two-photon signal at $750 \, \mbox{GeV}$ at the LHC seemed to indicate unambiguously decay of a scalar analog of the Higgs. If this were the case, the hyperpion mass in this model would have to be sufficiently small $\sim 10 ^ 2 \, \mbox{GeV}$ due to the direct connection between $\tilde{\pi} $ and $\tilde{\sigma}$ masses. The condition of small mixing of scalars $H$ and $\tilde{\sigma}$ in the conformal limit is $m_{\tilde{\sigma}}\approx \sqrt{3}m_{\tilde{\pi}}$~\cite{Pasechnik:2013bxa} and it means, in fact, that  $\tilde{\sigma}$ is a pNG boson of conformal symmetry. Then, it should be close in mass to other pNG states. In this case, signals of formation and decays of charged and neutral (stable!) H-pions would be observed at the collider~\cite{Lebiedowicz:2013fta,Beylin:2016kga}. Nature, however, turned out to be more sophisticated.

To consider the phenomenological manifestations, we postulate a certain hierarchy of scales for numerous degrees of freedom in the model. Namely, the pNG bosons are the lightest in the spectrum of possible hyperhadrons, and the triplet of H-pions are the lightest states of pNG. This arrangement of the scales follows from the assumption that the apparent violation of the symmetry SU(4) is a small perturbation by analogy with the violation of the dynamic symmetry in the orthodox QCD scheme.  There, the chiral symmetry is broken on a scale much larger than the mass scale of light quarks.

In the absence of new physics data from the LHC, we can use an estimate obtained on the assumption that the stable states in the model are dark matter candidates. In particular, the neutral H-pion $\tilde{\pi}^ 0$ and neutral hyperbaryon, $B^0$, can be such candidates. In this case, the analysis of the relic concentration of the dark matter makes it possible to estimate the range of masses of these particles.
Thus, there is a natural mutual influence and collaboration of astrophysical and collider studies. So, in this scenario of the Standard Model extension, it becomes possible to identify DM particles with two representatives of the pNG states. For quantitative analysis, however, a more accurate consideration of the mass spectrum of the H-pion triplet and mass splitting between $\tilde {\pi} ^ 0$ and hyperbaryon ${B^0}$ is necessary.

As for the mass splitting in the H-pion triplet, this parameter is defined by purely electroweak contributions~\cite{San08,doi:10.1142/S0217751X17500427} and is as follows:
\begin{equation}
\begin{aligned}\label{3.1.1}
\Delta m_{\tilde{\pi}} = {}&\dfrac{G_F M_W^4}{2\sqrt{2} \pi^2 m_{\tilde{\pi}}} \Biggl[ \ln \dfrac{M_Z^2}{M_W^2} -\beta_Z^2 \ln \mu_Z+ \beta_W^2 \ln \mu_W \\
&-\dfrac{4 \beta_Z^3}{\sqrt{\mu_Z}}\left ( \arctan \dfrac{2-\mu_Z}{2\sqrt{\mu_z}\beta_Z} +\arctan \dfrac{\sqrt{\mu_Z}}{2\beta_Z} \right)\\
&+\dfrac{4 \beta_W^3}{\sqrt{\mu_W}}\left( \arctan \dfrac{2-\mu_W}{2\sqrt{\mu_W}\beta_W} +\arctan \dfrac{\sqrt{\mu_W}}{2\beta_W} \right) \Biggr].
\end{aligned}
\end{equation}

Here, $\mu_V=M_V^2/m_{\tilde{\pi}}^2,\quad \beta_V=\sqrt{1-\mu_V/4}$, and $G_F$ denotes Fermi's constant. 
Taking the H-pion mass in a wide range $200\text{--}1500 \, \mbox{GeV}$,
from (\ref{3.1.1}) we found the value $\Delta m_{\tilde{\pi}} \approx 0.162\text{--}0.170 \, \text{GeV}$. 

Indeed, this small, non-zero and almost constant splitting of the mass in the triplet of the hyperpions obviously violates isotopic invariance. But at the same time, HG-parity remains a conserved quantum number. The reason is that the HG-parity is associated with a discrete symmetry, and not with a continuous transformation of the H-pion states. It is important to note that the inclusion of higher order corrections cannot destabilize a neutral weakly interacting H-pion, which is the lightest state in this pseudoscalar triplet. But charged H-pion states should decay by several channels. 

 In the strong channel, the width of the charged H-pion decay~\cite{Beylin:2016kga} can be written as
\begin{equation}\label{3.1.2}
\Gamma(\tilde{\pi}^{\pm}\to\tilde{\pi}^0 \pi^{\pm})=\dfrac{G^2_F}{\pi}f^2_{\pi}|U_{ud}|^2 m_{\tilde{\pi}}^{\pm} (\Delta m_{\tilde{\pi}})^2\bar\lambda(m^2_{\pi^\pm},m_{\tilde{\pi}^0}
^2;m_{\tilde{\pi}^{\pm}}^2).
\end{equation}

Here, $f_{\pi} =132 \,\mbox{MeV}$ and $\pi^{\pm}$ denotes a standard pion. The reduced triangle function is defined as
\begin{equation}\label{3.1.3}
\bar\lambda(a,b;c)=\left[1-2\dfrac{a+b}{c}+\dfrac{(a-b)^2}{c^2}\right]^{1/2}.
\end{equation}

For the decay in the lepton channel we get:
\begin{equation}\label{3.1.4}
\Gamma(\tilde{\pi}^{\pm}\to\tilde{\pi}^0 l^{\pm}\nu_l)=\dfrac{G^2_F
    m_{\tilde{\pi}^{\pm}}^3}{24\pi^3}\int_{q^2_1}^{q^2_2}\bar\lambda(q^2,m_{\tilde{\pi}^0}^2;m_{\tilde{\pi}^{\pm}}^2)^{3/2}
\left(1-\dfrac{3m^2_l}{2q^2}+\dfrac{m^6_l}{2q^6}\right)\,dq^2,
\end{equation}
where $q^2_1=m^2_l$, $q^2_2=(\Delta m_{\tilde{\pi}})^2$, and $m_l$ is a lepton mass.

Now, we can estimate decay widths of the charged H-pion and, correspondingly, lifetimes, and track lengths in these channels. To do this, we use \eqref{3.1.3}, \eqref{3.1.4}, and $\Delta m_{\tilde{\pi}}$ from \eqref{3.1.1} and get
\begin{equation}
\begin{aligned}\label{3.1.5}
\Gamma(\tilde{\pi}^{\pm}\to\tilde{\pi}^0 \pi^{\pm})&=6\cdot
10^{-17}\,\mbox{GeV},\,\,\,\tau_{\pi}=1.1\cdot10^{-8}\,\mbox{s},\,\,\,c\tau_{\pi}\approx 330\,\mbox{cm};\\
\Gamma(\tilde{\pi}^{\pm}\to\tilde{\pi}^0 l^{\pm}\nu_l)&=3\cdot10^{-15}\,\mbox{GeV},\,\,\,\tau_l=2.2\cdot10^{-10}\,\mbox{s},\,\,\,c\tau_l\approx 6.6\,\mbox{cm}.
\end{aligned}
\end{equation}

Then, at TeV scale, characteristic manifestations of H-pions can be observed in the Drell--Yan type reactions due to the following fingerprints:
\begin{enumerate}
\item large $E_{T,mis}$ reaction due to production of stable $\tilde \pi^0$ and neutrino from $\tilde \pi^{\pm}$ and/or $W^{\pm}$ decays, or two leptons from charged H-pion and $W$ decays (this is reaction of associated production, $W,\, \tilde \pi^{\pm},\, \tilde\pi^0$ final state of the process);

\item large $E_{T,mis}$ due to creation of two stable $\tilde \pi^0$ and neutrino from decay of charged H-pions, $\tilde \pi^{\pm}$, one lepton from $\tilde \pi^{\pm}$ and two quark jets from $W^{\pm}$ decay (the same final state with particles $W, \,\tilde \pi^{\pm}, \, \tilde\pi^0$);

\item large $E_{T,mis}$ due to two stable neutral H-pions and neutrino from charged
H-pion decay, two leptons (virtual $Z$, and $\tilde \pi^+, \, \tilde\pi^-$ in the final state);

\item large $E_{T,mis}$ due to two final neutral H-pions and neutrino from
$\tilde \pi^{\pm}$, one lepton which originated from virtual $W$, $\tilde \pi^+,\,\tilde\pi^0$ final states.
\end{enumerate}
Besides, H-pion signals can be seen due to two tagged jets in vector-boson-fusion channel in addition to main characteristics of the stable hyperpion---$E_{mis}\sim m_{\tilde \pi}$ and accompanying leptons.

Obviously, targeted search for such signals is possible only when we know, at least approximately, the range of hyperpion mass values. These estimates can be obtained by calculating the relic content of hidden mass in the Universe and comparing it with recent astrophysical data. Within the framework of the model, such calculations were made (see below). The possible values of the H-pion mass are in the range from $600\text{--}700~\mbox{GeV}$ to $1200\text{--}1400~\mbox{GeV}$, while the naturally $\tilde \sigma$-meson is quite heavy---we recall that its mass is directly related to the masses of H-pions in the case of small $H$-$\tilde \sigma$ mixing. Cross sections of the reactions above (with large missed energy and momentum) are too small to be detected without special and careful analysis of specific events with predicted signature. The number of these events is also evidently small, and the signal can be hardly extracted from the background because there is a lot of events with decaying $W$-bosons and, correspondingly, neutrino or quark jets. 

So, another interesting process to probe into the model of this type is the production and decay of a scalar H-meson $\tilde{\sigma}$ at the LHC; this production is possible at the tree level, however, the cross section is strongly damped due to small mixing. A small value of the mixing angle, $\theta_s$, suppresses the cross section by an extra multiplier $\sin^2 \theta_s$ in comparison with the standard Higgs boson production.

However, at one-loop level it is possible to get single
and double H-sigma in the processes of vector-vector fusion. Namely, in $V^{*}V'^{*}\to \tilde{\sigma}, \,2\tilde{\sigma}$ and/or in the decay through hyperquark triangle loop $\Delta$, i.e., $V^{*}\to \Delta \to V^{'}\tilde{\sigma},\,2\tilde{\sigma}$.
Here, $V^{*}$ and $V^{'}$ are intermediate or final vector bosons.

Now, a heavy H-sigma can decay via loops of hyperquarks or hyperpions $\tilde{\sigma}\to V_1V_2$, where $V_{1,2}=\gamma,Z,W$. Besides, the main decay modes of H-sigma are  $\tilde{\sigma}\to
\tilde{\pi}^0 \tilde{\pi}^0,\,\tilde{\pi}^+\tilde{\pi}^-$; these are described by tree-level diagrams that predict large decay width for $m_{\tilde{\sigma}}\geqslant 2m_{\tilde{\pi}}$. As we will see below (from the DM relic abundance analysis), at some values of $\tilde \pi$ and $\tilde \sigma$ masses these channels are opened. In the small mixing limit, the width is
\begin{equation}\label{3.1.6}
\Gamma(\tilde{\sigma}\to\tilde{\pi}\tilde{\pi})=\dfrac{3u^2\lambda^2_{11}}{8\pi
    m_{\tilde{\sigma}}}\left(1-\dfrac{4m^2_{\tilde{\pi}}}{m^2_{\tilde{\sigma}}}\right),
\end{equation}
and it depends strongly on the parameter $\lambda_{11}$.

An initial analysis of the model parameters was carried out in~\cite{Pasechnik:2013bxa}, using the value $\lambda_{11}$ (it is denoted there as $\lambda_{HC}$) and $u$, from (\ref{3.1.6}) we get:
$\Gamma(\tilde{\sigma}\to\tilde{\pi}\tilde{\pi})\gtrsim 10\,\mbox{GeV}$ when $m_{\tilde{\sigma}}\gtrsim 2m_{\tilde{\pi}}$.

The smallness of $H$--$\tilde{\sigma}$ mixing, as it is dictated by conformal approximation, results in the multiplier $\sin^2 \theta_s$ for all tree-level squared amplitudes for decay widths. Then, for $\tilde \sigma$ decay widths we have
\begin{equation}
\begin{aligned}\label{3.1.7}
\Gamma(\tilde{\sigma}\to f\bar{f})=&\dfrac{g_W^2 \sin^2 \theta_s}{32\pi}m_{\tilde{\sigma}}\dfrac{m^2_f}{M^2_W}(1-4\dfrac{m^2_f}{m^2_{\tilde{\sigma}}})^{3/2},\\
\Gamma(\tilde{\sigma}\to ZZ)=&\dfrac{g_W^2 \sin^2 \theta_s}{16\pi
c^2_W}\dfrac{M^2_Z}{m_{\tilde{\sigma}}}(1-4\dfrac{m^2_Z}{m^2_{\tilde{\sigma}}})^{1/2}[1+\dfrac{(m^2_{\tilde{\sigma}}-2M^2_Z)^2}{8M^4_Z}],\\
\Gamma(\tilde{\sigma}\to W^+W^-)=&\dfrac{g_W^2 \sin^2 \theta_s}{8\pi}\dfrac{M^2_W}{m_{\tilde{\sigma}}}(1-4\dfrac{m^2_W}{m^2_{\tilde{\sigma}}})^{1/2}
[1+\dfrac{(m^2_{\tilde{\sigma}}-2M^2_W)^2}{8M^4_W}].
\end{aligned}
\end{equation}

Here, $m_f$ is a mass of standard fermion $f$ and $c_W=\cos \theta_W$. 

Recall that the two-photon decay of the Higgs boson is the very main channel in which the deviation of the experimental data from the predictions of the SM was originally found. Analogically, we consider a $\tilde{\sigma}\to \gamma\gamma$ decay which occurs through loops of heavy hyperquarks and H-pions; the width has the following form:
\begin{equation}\label{3.1.8}
\Gamma(\tilde{\sigma}\to \gamma\gamma)=\dfrac{\alpha^2m_{\tilde{\sigma}}}{16\pi^3
    } |F_Q+F_{\tilde{\pi}}+F_{\tilde{a}}+F_W+F_\text{top}|^2.
\end{equation}

Here, $F_Q$, $F_{\tilde{\pi}}$, $F_W$, and $F_\text{top}$ are contributions from the H-quark, H-pion, $W$-boson, and top-quark loops; they can be presented as follows:
\begin{equation}
\begin{aligned}\label{3.1.9}
F_Q&=-2\kappa \dfrac{M_Q}{m_{\tilde{\sigma}}}[1+(1-\tau^{-1}_Q)f(\tau_Q)],  \\
F_{\tilde{\pi}}&=\dfrac{g_{\tilde{\pi}\tilde{\sigma}} }{m_{\tilde{\sigma}}}[1-\tau^{-1}_{\tilde{\pi}}f(\tau_{\tilde{\pi}})],\,\,\,g_{\tilde{\pi}\tilde{\sigma}}\approx u\lambda_{11},  \\
F_{\tilde{a}}&=\dfrac{g_{\tilde{a}\tilde{\sigma}}}{m_{\tilde{\sigma}}}[1-\tau^{-1}_{\tilde{a}}f(\tau_{\tilde{a}})],\,\,\,g_{\tilde{a}\tilde{\sigma}}\approx u\lambda_{12},  \\
F_W&=-\dfrac{g_W\sin\theta_s m_{\tilde{\sigma}}}{8M_W}
[2+3 \tau_W^{-1} + 3 \tau_W^{-1}(2-\tau_W^{-1}) f(\tau_W)],  \\
F_\text{top}&=\dfrac{4}{3}\dfrac{g_W\sin \theta_s M^2_t}{m_{\tilde{\sigma}}M_W} [1+(1-\tau^{-1}_t)f(\tau_t)],
\end{aligned}
\end{equation}
and
\begin{equation}
\begin{aligned}\label{3.1.10}
f(\tau)=&\arcsin^2\sqrt{\tau},\,\,\,\tau<1,\\
f(\tau)=&-\dfrac{1}{4}\Biggl[\ln\dfrac{1+\sqrt{1-\tau^{-1}}}{1-\sqrt{1-\tau^{-1}}}-i\pi\Biggr]^2,\,\,\,\tau
>1.
\end{aligned}
\end{equation}

As it is seen, contributions from $W$- and $t$-quark loops are induced by non-zero $\tilde{\sigma}\text{--}H$ mixing. Taking necessary parameters from~\cite{Pasechnik:2013bxa}, the width is evaluated as $\Gamma(\tilde{\sigma}\to \gamma\gamma) \approx 5\text{--}10 \,\, \mbox{MeV}$.

Obviously, the process $p\bar{p}\to \tilde\sigma \to \text{all}$ should be analyzed quantitatively after integration of cross section of quark subprocess with partonic distribution functions. It is reasonable, however, to get an approximate value of the vector boson fusion cross section $VV\to\tilde\sigma(s)\to \text{all}$, $V=\gamma, Z, W$. 

The useful procedure to calculate the cross section with a suitable accuracy is the method of factorization~\cite{Kuk_Vol}; this approach is simple and for the cross section estimation it suggests a clear recipe:
\begin{equation}\label{3.1.11}
\sigma(VV\to\tilde{\sigma}(s))=\dfrac{16\pi^2\Gamma(\tilde{\sigma}(s)\to VV)}{9\sqrt{s}\,\bar{\lambda}^2(M_V^2,M_V^2;s)}\,\rho_{\tilde{\sigma}}(s),
\end{equation}
where $\tilde{\sigma}(s)$ is $\tilde{\sigma}$ in the intermediate state having energy $\sqrt{s}$. A partial decay width is denoted as $\Gamma(\tilde{\sigma}(s)\to VV)$. The density of probability,
$\rho_{\tilde{\sigma}}(s),$ can be written as
\begin{equation}\label{3.1.12}
\rho_{\tilde{\sigma}}(s)=\dfrac{1}{\pi}\dfrac{\sqrt{s}\,\Gamma_{\tilde{\sigma}}(s)}{(s-M^2_{\tilde{\sigma}})^2+s\,\Gamma^2_{\tilde{\sigma}}(s)}.
\end{equation}

Here, $\Gamma_{\tilde{\sigma}}(s)$ is the total width of virtual $\tilde{\sigma}$-meson having $M_{\tilde{\sigma}}=\sqrt{s}$. At this energy we get exclusive cross section changing the numerator in (\ref{3.1.12}), namely $\Gamma_{\tilde{\sigma}}\,\to\,\Gamma(\tilde{\sigma}\to V^{'}
V^{'})=\Gamma_{\tilde{\sigma}}\cdot Br(\tilde{\sigma}\to V^{'} V^{'})$; for the cross section now we have
\begin{align}\label{3.1.13}
\sigma(VV\to\tilde{\sigma}\to V^{'} V^{'})=&\dfrac{16\pi}{9}\dfrac{Br(\tilde{\sigma}\to VV)Br(\tilde{\sigma}\to V^{'}
V^{'})}{m^2_{\tilde{\sigma}}(1-4M^2_V/m^2_{\tilde{\sigma}})}\notag\\&\approx\dfrac{16\pi}{9 m^2_{\tilde{\sigma}}}\cdot Br(\tilde{\sigma}\to VV)Br(\tilde{\sigma}\to V^{'}V^{'}).
\end{align}

Now, when $M^2_{\tilde{\sigma}}\gg M^2_V$ the cross section considered is determined by the branchings of H-sigma decay and the value of $M_{\tilde{\sigma}}$. Note, if $2m_{\tilde{\pi}}>M_{\tilde{\sigma}}$ H-sigma dominantly decays through following channels $\tilde{\sigma}\to WW, ZZ$. In this case, we get for $\tilde \sigma$ a narrow peak $(\Gamma\lesssim 10\text{--}100\, \mbox{MeV}$). 

As we said earlier, up to the present, there are no signals from the LHC about the existence of a heavy scalar state that mixes with the Higgs boson. Therefore, we are forced to estimate the mass of the H-sigma relying on astrophysical data on the DM concentration. Namely, we can consider H-pions as stable dark matter particles and then take into account the connection of their mass with the mass of $\tilde \sigma$-meson in the (almost) conformal limit.

Now, we should use the cross section which is averaged over energy resolution. As a result, the value of the cross section is reduced significantly. More exactly, for $2m_{\tilde{\pi}}<M_{\tilde{\sigma}}$ the dominant channel is $\tilde{\sigma}\to \tilde{\pi}\tilde{\pi}$ with a wide peak $(\Gamma\sim 10\,
\mbox{GeV})$. So, $Br(\tilde{\sigma}\to VV)$ is small and consequently the cross section of H-meson prodution is estimated as very small. 

Thus, with a sufficiently heavy (with mass $(2-3)\, \mbox{TeV}$) second scalar meson, the main fingerprint of its emergence in the reaction is a wide peak induced by the strong decay $\tilde{\sigma}\to 2\tilde{\pi}$. It is accompanied by final states with two photons, leptons and quark jets originating from decays of $WW$, $ZZ$, and standard $\pi^{\pm}$. Besides, it occurs with some specific decay mode of $\tilde \sigma$ with two final stable $\tilde\pi^0$. This channel is specified by a large missed energy; charged final H-pions result in a signature with missed energy plus charged leptons. 

As it was shown, existence of global U(1) hyperbaryon symmetry leads to the stability of the
lightest neutral H-diquark. In the scenario considered, we suppose that charged H-diquark states decay to the neutral stable one and some other particles. Moreover, we also assume that these charged H-diquarks are sufficiently heavy, so their contributions into processes at $(1-2) \, \mbox{TeV}$ are negligible~\cite{Beylin:2016kga}.

Thus, having two different stable states---neutral H-pion and the lightest H-diquark with conserved H-baryon number---we can study a possibility to construct dark matter from these particles. This scenario with two-component dark matter is an immanent consequence of symmetry of this type of SM extension. Certainly, emergence of a set of pNG states together with heavy hyperhadrons needs careful and detailed analysis. At first stage, the mass splitting between stable components of the DM, not only H-pions, should be considered. Importantly, the model does not contain stable H-baryon participating into electroweak interactions. It means that any constraints for the DM relic concentration are absent for this case.

Note, $\tilde \pi^0$ and $B^0$ have the same tree level masses, so it is important to analyze the mass splitting $\Delta M_{B\tilde \pi} = m_{B^0}-m_{\tilde{\pi}^0}$. As it follows from calculations, this parameter depends on electroweak contributions only, all other (strong) diagrams are canceled mutually. Then, we get:
\begin{equation}
\begin{aligned} \label{3.1.14}
\Delta M_{B\tilde \pi}=&\dfrac{-g^2_2m_{\tilde{\pi}}}{16\pi^2}\Bigg[ 8\beta^2-1-(4\beta^2-1)\ln\dfrac{m_{\tilde{\pi}}^2}{\mu^2}+2\dfrac{M_W^2}{m_{\tilde{\pi}}^2}\left(\ln\dfrac{M_{W}^2}{\mu^2}-\beta^2 \ln\dfrac{M_W^2}{m_{\tilde{\pi}}^2}  \right)
 \\ 
&-8\dfrac{M_W}{m_{\tilde{\pi}}}\beta^3\left(\arctan\dfrac{M_W}{2m_{\tilde{\pi}}\beta}+\arctan\dfrac{2m_{\tilde{\pi}}^2-M_W^2}{2m_{\tilde{\pi}}M_W\beta} \right)
\bigg],
\end{aligned}
\end{equation}
where $\beta=\displaystyle \sqrt{1-\dfrac{M_W^2}{4m_{\tilde{\pi}}^2}}$. An important point is that $\Delta M_{B\tilde \pi}$ dependence on a renormalization point results from the coupling of the pNG states with H-quark currents of different structure at close but not the same energy scales. Thus, the dependence of the characteristics of the DM on the renormalization parameter is necessarily considered  when analyzing the features of the DM model.

We remind that it is assumed that not-pNG H-hadrons (possible vector H-mesons, etc.) manifest itself at much more larger energies. It results from the smallness of the scale of explicit SU(4) symmetry breaking comparing with the scale of dynamical symmetry breaking. This hierarchy of scales copies the QCD construction. 

Now, since the effects of hyperparticles at the collider are small, and an interesting scenario of a two-component DM arises, let us consider in more detail the possibility of describing the dark matter candidates in the framework of this model~\cite{Conf_1,IJMPA_2019}. At the same time, we should note the importance of previous studies of the DM scenarios based on vector-like technicolor in the papers ref.~\cite{Pasechnik:2013kya,Pasechnik:2014ida,doi:10.1142/S0217751X17500427,Pasechnik:2013bxa, TCDM5,TCDM6,Gudnason:2006yj,Cirelli}. Several quite optimistic versions of the DM description (including technineutrons, B-baryons, etc.) were considered, which, however, did not have a continuation, since they relied on a number of not quite reasonable assumptions---in particular, that B-baryons form~the~triplet.

When we turn to the hypercolor model, it will be necessary not only to calculate the total annihilation cross section for dark matter states, but to analyze the entire kinetics of freezing out of DM particles.  The reason is that the mass splittings in the H-pions multiplet  and between masses of two components are small (in the last case this parameter can be suggested as small). Then, the coupled system of five Boltzmann kinetic equations should be solved. Namely, two states of the neutral H-baryon, $B^0, \, \, \bar B ^ 0$, neutral H-pion and also two charged H-pions should be considered. Such cumbersome kinetics are a consequence of proximity of masses of all particles participating in the process of formation of residual DM relic concentration. So, the co-annihilation processes~\cite{GriestSeckel} can contribute significantly to the cross section of annihilation. It had been shown also in previous vector-like scenario~\cite{doi:10.1142/S0217751X17500427}. 

Now, for each component of the DM and co-annihilating particles, $i,j = \tilde{\pi}^+,\,\tilde{\pi}^-,\tilde{\pi}^0;\mu, \nu =B,\, \bar{B}$, we have the basic Boltzmann equation (\ref{Boltzman1a})--(\ref{Boltzman1b}) (neglecting reactions of type $iX\leftrightarrow jX$):    
\begin{equation} 
\begin{aligned} 
\label{Boltzman1a}
\dfrac{da^3n_i}{a^3dt}=-&\sum\limits_j<\sigma v>_{ij}\left(n_in_j-n_i^{eq}n_j^{eq}\right)-\sum\limits_j\Gamma_{ij}\left(n_i-n_i^{eq}\right)-
 \\
&\sum\limits_{j,\mu,\nu}<\sigma v>_{ij\rightarrow \mu\nu}\left(n_in_j-\dfrac{n_i^{eq}n_j^{eq}}{n_{\mu}^{eq}n_{\nu}^{eq}}n_{\mu}n_{\nu}\right)+  \\
&\sum\limits_{j,\mu,\nu}<\sigma v>_{\mu\nu\rightarrow ij}\left(n_{\mu}n_{\nu}-\dfrac{n_{\mu}^{eq}n_{\nu}^{eq}}{n_{i}^{eq}n_{j}^{eq}}n_{i}n_{j}\right).
\end{aligned}
\end{equation}

We also get
\begin{equation} 
\begin{aligned}
\label{Boltzman1b}
\dfrac{da^3n_{\mu}}{a^3dt}=-&\sum\limits_{\nu}<\sigma v>_{\mu\nu}\left(n_{\mu}n_{\nu}-n_{\mu}^{eq}n_{\nu}^{eq}\right)+  \\
&\sum\limits_{\nu,i,j}<\sigma v>_{ij\rightarrow \mu\nu}\left(n_in_j-\dfrac{n_i^{eq}n_j^{eq}}{n_{\mu}^{eq}n_{\nu}^{eq}}n_{\mu}n_{\nu}\right)- \\
&\sum\limits_{\nu,i,j}<\sigma v>_{\mu\nu\rightarrow ij}\left(n_{\mu}n_{\nu}-\dfrac{n_{\mu}^{eq}n_{\nu}^{eq}}{n_{i}^{eq}n_{j}^{eq}}n_{i}n_{j}\right),
\end{aligned}
\end{equation}
where %we denote for $i,j$ and also for $\mu, \nu$ components: 
\begin{equation} 
\begin{aligned}
<\sigma v>_{ij}=&<\sigma v>(ij\rightarrow XX),  \\
<\sigma v>_{ij\rightarrow \mu\nu}=&<\sigma v>(ij\rightarrow \mu\nu),  \\
\Gamma_{ij}=& \Gamma(i\rightarrow jXX).
\end{aligned}
\end{equation}

Because of decays of charged H-pions, the main parameters in this calculation are total densities of $\tilde \pi$, $B^0$ and $\bar{B^0}$, namely, $n_{\tilde \pi}=\sum_{i}n_i$ and $n_{B}=\sum_{\mu}n_{\mu}$. 
Using an equilibrium density, $n_{eq}$, for describing co-annihilation, we estimate $n_i/n \approx n_i^{eq}/n^{eq}$. Then, the system of equations can be rewritten as 
\begin{equation} 
\begin{aligned}
\dfrac{da^3n_{\pi}}{a^3dt}=&\bar{<\sigma v>}_{\tilde \pi}\left(n_{\tilde \pi}^2-\left(n_{\tilde \pi}^{eq}\right)^2\right)
-<\sigma v>_{\tilde \pi \tilde \pi}\left(n_{\tilde \pi}^2-\dfrac{\left(n_{\tilde \pi}^{eq}\right)^2}{\left(n_{B}^{eq}\right)^2}n_{B}^2\right)+  \\
&<\sigma v>_{BB}\left(n_{B}^2-\dfrac{\left(n_{B}^{eq}\right)^2}{\left(n_{\tilde \pi}^{eq}\right)^2}n_{\tilde \pi}^2\right),\label{Boltzman2a}
\end{aligned}
\end{equation}
\begin{equation} 
\begin{aligned}
\dfrac{da^3n_{B}}{a^3dt}=&\bar{<\sigma v>}_{B}\left(n_{B}^2-\left(n_{B}^{eq}\right)^2\right)
+<\sigma v>_{\tilde \pi \tilde \pi}\left(n_{\tilde \pi}^2-\dfrac{\left(n_{\tilde \pi}^{eq}\right)^2}{\left(n_{B}^{eq}\right)^2}n_{B}^2\right)-  \\
&<\sigma v>_{BB}\left(n_{B}^2-\dfrac{\left(n_{B}^{eq}\right)^2}{\left(n_{\tilde \pi}^{eq}\right)^2}n_{\tilde \pi}^2\right),\label{Boltzman2b}
\end{aligned}
\end{equation}
where
\begin{equation} 
\begin{aligned}
\bar{<\sigma v>}_{\tilde \pi}=&\dfrac{1}{9}\sum\limits_{i,j}<\sigma v>_{ij},~~~~\bar{<\sigma v>}_{B}=\dfrac{1}{4}\sum\limits_{\mu,\nu}<\sigma v>_{\mu\nu},
 \\
<\sigma v>_{\tilde \pi \tilde \pi}=&\dfrac{1}{9}(<\sigma v>(\tilde \pi^0 \tilde \pi^0\rightarrow B\bar{B})+2<\sigma v>(\tilde \pi^+\tilde \pi^-\rightarrow B\bar{B})),
 \\
<\sigma v>_{BB}=&\dfrac{1}{2}(<\sigma v>(B\bar{B}\rightarrow\tilde \pi^0 \tilde \pi^0)+<\sigma v>(B\bar{B}\rightarrow\tilde \pi^-\tilde \pi^+)).
\end{aligned}
\end{equation}

Now, it is reasonable to consider the ratio $m_{\tilde \pi }/M_B \approx 1$ using a suitable value of the renormalization parameter in the mass splitting between $m_{\tilde \pi^0}$ and $M_{B^0}$, more exactly, $\Delta M_{B^0\tilde \pi^0}/m_{\tilde \pi^0} \lesssim 0.02$. Then, the~system of kinetic equations simplifies further. Having $n_{B}^{eq}/n_{\tilde \pi}^{eq}=2/3$, we come to the following form of equations:
\begin{equation} 
\begin{aligned}
\dfrac{da^3n_{\tilde \pi}}{a^3dt}=&\bar{<\sigma v>}_{\tilde \pi}\left(n_{\tilde \pi}^2-\left(n_{\tilde \pi}^{eq}\right)^2\right)
-<\sigma v>_{\tilde \pi \tilde \pi}\left(n_{\tilde \pi}^2-\dfrac{9}{4}n_{B}^2\right)+  \\
&<\sigma v>_{BB}\left(n_{B}^2-\dfrac{4}{9}n_{\tilde \pi}^2\right),\label{Boltzman3a}
\end{aligned}
\end{equation}
\begin{equation} 
\begin{aligned}
\dfrac{da^3n_{B}}{a^3dt}=&\bar{<\sigma v>}_{B}\left(n_{B}^2-\left(n_{B}^{eq}\right)^2\right)
+<\sigma v>_{\tilde \pi \tilde \pi}\left(n_{\tilde \pi}^2-\dfrac{9}{4}n_{B}^2\right)-  \\
&<\sigma v>_{BB}\left(n_{B}^2-\dfrac{4}{9}n_{\tilde \pi}^2\right).
\label{Boltzman3b}
\end{aligned}
\end{equation}

 When masses of the DM components are close to each other, it is necessary to take into account the temperature dependence~\cite{GriestSeckel} in the cross sections $<\sigma v>_{\tilde \pi \tilde \pi}$ and $<\sigma v>_{BB}$ of the processes: 
\begin{eqnarray} 
<\sigma v>_{BB}\approx <(a+bv^2)v>=\dfrac{2}{\sqrt{\pi}x}\left(a+\dfrac{8b}{x}\right),
\end{eqnarray}
where $x=m_{\tilde \pi}/T$, $v$ is the relative velocity of final particles. 

There are commonly used notations, which are convenient for  solve the system, $Y=n/s$ and $x=m_{\tilde \pi}/T$, where $s$ is the density of entropy. Then, neglecting small terms $~\Delta M_{\tilde \pi}/M_{B}$ we have:
\begin{eqnarray} 
\label{Boltzman4a}
\dfrac{dY_{\pi}}{dx}=g(x,T)\left[\lambda_{\tilde \pi}((Y_{\tilde \pi}^{eq})^2-Y^2_{\tilde \pi})-\lambda_{\tilde \pi \tilde \pi}\left(Y_{\tilde \pi}^2-\dfrac{9}{4}Y_B^2\right)+ \lambda_{BB}\left(Y_{B}^2-\dfrac{4}{9}Y_{\tilde \pi}^2\right)\right],
\\
\label{Boltzman4b}
\dfrac{dY_{B}}{dx}= g(x,T)\left[\lambda_{B}((Y_{B}^{eq})^2-Y^2_{B})+\lambda_{\tilde \pi \tilde \pi}\left(Y_{\tilde \pi}^2-\dfrac{9}{4}Y_B^2\right)- \lambda_{BB}\left(Y_{B}^2-\dfrac{4}{9}Y_{\tilde \pi}^2\right)\right].
\end{eqnarray}

The energy density is determined by a set of relativistic degrees of freedom, this function can be written in a convenient form as
\begin{equation} 
\begin{array}{c}
g(x,T)=\displaystyle \dfrac{\sqrt{g(T)}}{x^2}\left\{1+ \displaystyle \dfrac{1}{3}\dfrac{d(\log g(T))}{d(\log T)}\right\}  \\[1ex]
\simeq \dfrac{115}{2}+\dfrac{75}{2}\tanh\left[2.2\left(\log_{10}T+0.5\right)\right]+10\tanh\left[3\left(\log_{10}T-1.65\right)\right].
\end{array}
\end{equation}

Here, we use an approximated value of this parameter, which works in the numerical solution with a good accuracy~\cite{IJMPA_2019} and better than known approximation $g(T)\approx 100$; also, there are standard notations from~\cite{Steigman}:  $\lambda_i=2.76\times 10^{35}m_{\tilde \pi}<\sigma v>_{i}$, $Y_{\tilde \pi}^{eq}=0.145(3/g(T))x^{3/2}e^{-x},\, Y_{B}^{eq}=0.145(2/g(T))x^{3/2}e^{-x}$, 

The DM relic density, $\Omega h^2$, is expressed in terms of relic abundance and critical mass density, $\rho$ and $\rho_{crit}$:
\begin{eqnarray} 
\Omega h^2 = \dfrac{\rho}{\rho_{crit}}h^2=\dfrac{m s_0 Y_0 }{\rho_{crit}}h^2\simeq 0.3 \times 10^9 \dfrac{m}{\text{GeV}} Y_0.
\end{eqnarray}

Present time values are denoted by the subscript ``0''.

After the replacement $W=\log Y$~\cite{Steigman}, the system of kinetic equations is solved numerically. As it is shown in detail in~\cite{IJMPA_2019}, there is a set of regions in a plane of H-pion and H-sigma masses, where it is possible to get the value of the DM relic density in a correspondence with the modern astrophysical data. More exactly, the H-pion fraction is described by the following intervals: $0.1047 \le \Omega h^2_{HP}+\Omega h^2_{HB} \le 0.1228$ and $\Omega h^2_{HP}/(\Omega h^2_{HP}+\Omega h^2_{HB}) \le 0.25$). There are also some slightly different areas having all parameters nearly the same. However, H-pions make up just over a quarter of dark matter density, more exactly, $0.1047 \le \Omega h^2_{HP}+\Omega h^2_{HB} \le 0.1228$ and $0.25\le \Omega h^2_{HP}/(\Omega h^2_{HP}+\Omega h^2_{HB}) \le 0.4$.
Certainly, there are regions of parameters which are forbidden by restrictions by XENON collaboration~\cite{3,LUX_1,DARWIN_1}.

It is important that there are no regions of parameters where the H-pion component dominates in the dark matter density. The reason is that this hyperpion component interacts with vector bosons, $Z, \,W$,  at the tree level and, consequently, annihilates into ordinary particles much faster than stable $B^0$-baryons. The latter particles do not interact with standard vector bosons directly but only at loop level through H-quark and H-pion loops. It is a specific feature of SU(4) vector-like model having two stable pNG states. 

At this stage of analysis (without an account of loop contributions from $B^0-B^-$ annihilation), there are three allowable regions of parameters (masses):

{Area 1}: here $M_{\tilde{\sigma}}>2m_{\tilde \pi^0}$ and $u \ge M_{\tilde{\sigma}} $; at small mixing, $s_{\theta}\ll 1$, and large mass of H-pions we get a reasonable value of the relic density and a significant H-pion fraction; 

{Area 2}: here again $M_{\tilde{\sigma}}>2m_{\tilde \pi^0}$ and $u \ge M_{\tilde{\sigma}}$ but $m_{\tilde \pi}\approx 300-600 \,\, \mbox{GeV}$; H-pion fraction is small here, approximately, $(10-15)\%$;

{Area 3}: $M_{\tilde{\sigma}}<2m_{\tilde \pi}$---this region is possible for all values of parameters, but decay $\tilde \sigma \to \tilde \pi \tilde \pi$ is prohibited and two-photon signal from reaction $pp\to \tilde \sigma \to \gamma\gamma X$ would have to be visible at the LHC. Simultaneously, H-pion fraction can be sufficiently large, up to $40\%$ for large $m_{\tilde \pi^0}\sim 1 \,\mbox{TeV}$ and small angle of mixing. 

Thus, from kinetics of two components of hidden mass it follows that the mass of these particles can vary in the interval $(600\text{--}1000)\, \mbox{GeV}$ in agreement with recent data on the DM relic abundance. Having these values, it is possible to consider some manifestations of the hidden mass structure in~the~model. 

Particularly, the inelastic interactions of high-energy cosmic rays with the DM particles can be interesting for studying the hidden mass distribution using signals of energetic leptons (neutrino) or photons which are produced in this scattering process~\cite{IJMPA_2019}.

Cosmic ray electrons can interact with the H-pion component via a weak boson in the process \mbox{$e \tilde \pi^0 \to \nu_e \tilde \pi^-$}, then charged $\tilde \pi^-$ will decay. In the narrow-width approximation we get for the cross section: 
$\sigma(e \tilde \pi^0 \to \nu_e \tilde \pi^0 l \nu'_l ) \approx \sigma((e \tilde \pi^0 \to \nu_e \tilde \pi^-)\cdot Br( \tilde \pi^-\to \tilde \pi^0 l \nu'_l ),$ branchings of charged hyperpion decay channels are: $Br( \tilde \pi^-\to \tilde \pi^0 e \nu'_e) \approx 0.01$ and also $Br( \tilde \pi^-\to \tilde \pi^0 \pi^-) \approx 0.99$. 

Considering final charged hyperpion $\tilde \pi^-$ near its mass shell, standard light charged pion produces neutrino $e \nu_e$ and $\mu \nu_{\mu}$ with following probabilities: $\approx 1.2\times 10^{-6}$ and $\approx$0.999, correspondingly. 

Then, in this reaction, an energetic cosmic electron produces electronic neutrino  due to vertex $We \nu_e$, and soft secondary $e' \nu'_e$ or $\mu \nu_{\mu}$ arise from charged H-pion decays. Now, there are final states with $Br(\tilde \pi^0 \nu_e \mu'\nu'_{\mu})\approx 0.99$ and $Br(\tilde \pi^0 \nu_e e'\nu'_e)\approx 10^{-2}$. Obviously, we use here some simple estimations, they can be justified in the framework of the factorization approach~\cite{Kuk_Vol}. Characteristic values of H-pion mass which can be used for the analysis are, for example, $m_{\tilde \pi^0}=800\,\, \mbox{GeV}$ and $1200\,\, \mbox{GeV}$. 

As it results from calculations, at initial electron energies in the interval $E_e = (100\text{--}1000) \,\, \mbox{GeV}$ the cross section of the process decreases from $O(10) \,\,\mbox{nb}$ up to 
$O(0.1) \,\,\mbox{nb}$ having maximum at small angles between electron and the neutrino emitted, i.e., inelastic neutrino production occurs in the forward direction (for more detail see figures in Ref.~\cite{IJMPA_2019}). In this approximation, the energy of the neutrino produced is proportional to the energy of the incident electron and depends on the mass of the dark matter particle very weakly.
The neutrino flux is calculated by integrating of spectrum, $dN/dE_{\nu}$, this flux depends on H-pion mass very weakly. In the interval $(50-350)\, \mbox{GeV}$ it decreases most steeply,  and then, down to energies $\sim$1 \mbox{TeV} the fall is smoother. 

Certainly, integrating the spectrum $dN/dE_{\nu}$, we can estimate the number of neutrino landing on the surface of IceTop~\cite{IceCube_1,IceCube_2} which is approximately one squared kilometer. 

Even taking into account some coefficients to  amplify the DM density near the galaxy center for the  symmetric Einasto profile, we have found the number of such neutrino events per year as very small, $N_{\nu} =(6-7)$, in comparison with the corresponding number of events for neutrino with energies in the multi-TeV region. Note, the Einasto profile modified in such manner reproduces well the hidden mass density value near Galaxy center in concordance with other DM profiles~\cite{Cirelli_1}. In any case, such small number of neutrino events at IceCube does not allow to study this interaction of cosmic rays with the DM effectively. Any other DM profile gives practically the same estimation of number of events for neutrino with these energies. Indeed, cross section of $\nu N$ interaction for small neutrino energies $\sim (10^2 - 10^3) \, \mbox{GeV}$ is much lower than for neutrino energies $\sim (10^1-10^5)\, \mbox{TeV}$. Consequently, all parameters of the signal detected, particularly, deposition of energy, intensity of Cherenkov emission, are noticeably worse. Because of the absence of good statistics of neutrino events, it is practically impossible to measure the neutrino spectrum of the predicted form. Probability of neutrino detection can be estimated in the concept of an effective area of the detector~\cite{IceCube_2,Eff1,Eff2,Eff6,Eff7_Rev}). In our case this probability is small, $P=10^{-10}\text{--}10^{-8}$~\cite{IJMPA_2019}, so we need some additional factor that can increase the flux of neutrino substantially.

In principle, some factors amplifying these weak signals of cosmic electron scattering off the DM can be provided by inhomogeneities in the hidden mass distribution, i.e., so called clumps~\cite{Clumps_1,Clumps_2,Clumps_3,Clumps_4,Clumps_subhalo,Clumps_N}. The scattering of cosmic rays off clusters of very high density~\cite{Clumps_highdens_1} can result in amplifying neutrino flux substantially~\cite{Clumps_amplify_1,Clumps_amplify_2}. 

Though the scattering process suggested can be seen due to specific form of neutrino flux, the expected number of events is too small to be measured in experiments at modern neutrino observatories. The weakeness of the signal is also resulted from effective bremsstrahlung of electrons and the smallness of electron fraction in cosmic rays, $\sim$1\%. Therefore, they are not so good probe for the DM structure; only if there are sharply non-homogeneous spatial distribution of hidden mass, the signal of production of energetic neutrino by cosmic electrons can be detected. It is an important reason to study inelastic scattering of cosmic protons, because they are more energetic and have a much larger flux. 

Besides, an important information of the nature and profile of hidden mass should be manifested in a specific form of the DM annihilation gamma spectrum from clumps~\cite{Clumps_5,Clumps_6}. This signal can be significantly amplified due to increasing of density of hidden mass inside clumps, corresponding cross section depends on the squared density in contrary with the energy spectrum of final particles (neutrino, for example) which is resulted from scattering. In the last case, cross section is proportional to a first degree of the DM density.

Indeed, in the vector-like model with the two lowest in mass neutral stable states, one from these components does not participate in the scattering reaction with leptons, so the flux of final particles is diminished. There are, however, annihilation channels of both components into charged secondaries which emit photons. Some important contributions into this process describe so called virtual internal bremsshtrahlung (VIB). This part of photon spectrum containing information on the DM structure may be about 30\%. Consequently,  a feature of the DM structure in the model (particularly, existence of two components with different tree-level interactions) can result in  some characteristic form of the annihilation diffuse spectra. 

Introducing a parameter which determines H-pion fraction in the DM density,
 \begin{equation}
 \kappa=\dfrac{\Omega h^2_{HP}}{\Omega h^2_{HP}+\Omega h^2_{HB} },
 \end{equation}
full annihilation spectrum is written as
 \begin{align}
 \dfrac{d(\sigma v)}{d E_{\gamma}}E_{\gamma}^2=\kappa^2\dfrac{d(\sigma v_{\tilde{\pi}^0\tilde{\pi}^0})}{d E_{\gamma}}E_{\gamma}^2+(1-\kappa)^2\dfrac{d( \sigma v_{B^0\bar{B}^0})}{d E_{\gamma}}E_{\gamma}^2.
 \end{align}
 
 Contributions of the DM components to the total cross section of production of diffuse photons differ because of distinction in tree-level interaction with weak bosons. Annihilation of hyperpions into charged states (in particular, $W$-bosons) gives the most intensive part of diffuse photon flux. However, $B^0$-baryons can provide a significant fraction of this flux in some regions of the model parameters, namely, if the DM particles have mass  $\approx$600 \mbox{GeV}. It should be noted also that $\sigma$-meson mass affects the cross section value changing it noticeably: from $-10 \,\%$ to $+50\,\%$, approximately. This effect is seen better for the DM component mass $\approx$800 \mbox{GeV} when contribution from $B^0$ is not so prominent. Obviously, contributions to the gamma flux intensity from annihilation of different DM components contribute to the gamma flux in correspondence with the model content and structure. Thus, there appears a sign of the existence of two Dark matter components, observed in the form of a specific humped curve of the photon spectrum, due to virtual internal bremsstrahlung subprocesses. This effect should be considered in detail, because it is necessary to have much more astrophysics data together with an accurate analysis of all possible contributions to the spectrum for various regions of parameters. Certainly, because of high density of interacting particles, reactions of annihilation into photons in the DM clumps can be seen much better, and it also should be studied.
 
There is a set of possible observing consequences of the DM particle  origin, structure and interactions produced by vector-like extension of the SM. Some of them have been analyzed quantitatively, while the analysis of others is still in progress. As there are plenty of  additional heavy degrees of freedom in this model, they induce new effects that should be considered to predict observable and measurable phenomena. Moreover, the numerical estimations of model parameters and analysis of the effects above are based on some assumptions about spectrum of hyperhadrons. Particularly, it is suggested that charged di-hyperquark states, $B^{\pm}$, have masses which are much larger than neutral-state masses. It allows us to eliminate a lot of possible subprocesses with these particles and simplify substantially the system of kinetic equations for the DM components. This approach is quite reasonable. 

Extension of the vector-like model symmetry, from SU(4) to SU(6), unambiguously results in a much larger number of additional H-hadrons which spawn a great quantity of new processes and effects. As noted above, there is an invariance of the model physical Lagrangian with respect to some additional symmetries, as a result of which we obtain a number of stable states and it is necessary to study their possible manifestations.
Consideration of a new variant of the vector-like model of H-quarks is at the very beginning, therefore now we can define only some possible scenarios.

In the scenario with $Y = Y_S = 0$, two stable neutral states, H-pion, $B^0$ and also the lightest charged $B^{\pm}$ occur, as it is dictated by hyper-$G$-parity. In this case, we again have the opportunity to construct hidden mass from several components, as it was done in the previous version of SU(4) symmetry. However, a quantitative analysis of the mass difference for B-diquarks is necessary in order to assess the importance of the co-annihilation process for them. Assuming this mass splitting to be small, one can predict that the characteristics of a two-component DM in this scenario will not differ much from the previous version. Namely, we expect the masses of all dark matter components to be in the interval $(0.8\text{--}1.2)\, \mbox{TeV}$ providing corresponding DM density. 

Very interesting consequences follow from an occurrence of the stable charged state. First, the charged H-hadrons interact electromagnetically with cosmological plasma, so the hidden mass can be split from the plasma much later in comparison with the purely neutral DM. Second, there should be tree-level annihilation of these DM states into photons with an observable flux of specific form. Certainly, these conclusions make sense if relative concentration of the charged component is not small. Known data on the gamma spectrum from cosmic telescopes should help to establish necessary restrictions for the scenario parameters. 

Moreover, the stable charged H-hadron can be seen in the collider experiments at corresponding energies. These heavy particles in the final states and neutral stable particles should be observed in the characteristic events with large missed energy. The cross sections of reactions  and energies of detectable secondaries (hadronic jets and/or leptons) depend on the mass splitting between neutral and charged states. 

The charged stable H-hadron should also be prominent in the scattering off the nuclei in underground experiments, we can expect that the corresponding cross section will be larger then in the stable neutral component scattering due to exchanges via vector bosons, not only through intermediate scalar mesons. However, known restrictions for measurable cross sections which follow from   experiments at the underground setup will predict then more heavy stable particles in this model.

Note also that the next possible scenario with $Y_Q = 0$, $Y_S = \pm 1/2$ is less interesting for describing the DM properties because of the absence of stable states. (There is, possibly, a very special case with the one H-baryon state stable, the case should be considered separately, this work is in progress.)

Considering the physical Lagrangian for SU(6) vector-like extension, we find an important feature of the model: in this case there arise interactions of K-doublets and B-states with standard vector bosons (see the section above). These interactions can both amplify channels of new particles production at the collider and increase cross section of the DM annihilation and co-annihilation. Then, possible value of the DM components mass should also be larger to provide a suitable hidden mass density. In any case, these scenarios should be carefully analyzed before we can formulate a set of predictions for collider and astrophysical measurements. As it is seen, the vector-like extensions of the SM allow us to suggest some interesting scenarios with new stable heavy objects---H-hadrons---which can manifest itself  both in events with large missed energy at the LHC and in astrophysical signals such as spectrum of photons and/or leptons from various sources in the Galaxy.

\section{Dark Atom Physics and Cosmology}\label{darkatoms}
The approach of dark atoms, proposed and developed in~\cite{I,KK,FKS}, had followed the idea by Sh.L. Glashow~\cite{13d} on dark matter species as electromagnetically bound systems of new stable charged particles. The potential danger for this approach is the possibility of overproduction of anomalous hydrogen, being a bound state of a heavy +1 charged particle with ordinary electrons. Hence +1 charged particles should be unstable to avoid such an overproduction. Moreover, primordial heavy stable $-$1 charged particles, being in deficit relative to primordial helium are all captured by primordial helium nuclei, forming a +1 chareged ion, as soon as helium is produced in the course of Big Bang nucleosynthesis~\cite{FK}. Therefore charged dark atom constituents should have even charge, which is a double charge in the simplest case.

The abundance of particles with charge +2, bound with ordinary electrons, should be suppressed to satisfy the experimental upper limits on the anomalous helium. They should be either produced in deficit relative to the corresponding $-$2 charged particles~\cite{I,KK,KK2}, or there should be some special mechanism, suppressing the abundance of anomalous helium in the terrestrial matter~\cite{FKS}.

These constraints distinguish negative even charged particles as possible constituents of dark atoms. Particles with charge $-2$ are captured by primordial helium and form $O$-helium dark atom. Particles $X$ with even negative charge form $X$-nuclearites: the $-4$ charged capture two helium nuclei and form $X$-berillium ($X$Be), with the charge $-$6 capture three helium nuclei to form $X$-carbon ($X$C) and with the charge $-8$--$X$-oxygen ($X$O), in which four helium nuclei are bound with $-8$ charged particle. The~existing examples of $O$ or $X$ particles exhibit their leptonic or lepton-like nature, and the properties and effects of the coresponding dark atoms are determined by their nuclear-interacting helium shells. It naturally puts $O$He and X-nulearites in the list of hadronic dark matter candidates.

\subsection{Dark Atoms Structure, Effects and Probes}\label{DAstructure}
General analysis of the bound states of massive negatively charged  particle with~nuclei was proposed in~\cite{Cahn,Pospelov,Kohri}. It assumed a simplified description of nuclei as homogeneously charged spheres and that the charged particle doesn't possess strong interaction. The structure of the corresponding bound state depends on the value of parameter 
 $a = Z Z_o \alpha A m_p R$, where $Z$, \beq R \sim 1.2 A^{1/3} /(200 \MeV) \label{rA}\eeq and $A$ are, respectively, charge, radius and atomic number of the nucleus. In the Equation (\ref{rA}) $Z_o$ is the charge of particle, $\alpha$ is the fine structure constant and $m_p$ stands for the proton mass. For $0 < a < 1$ the bound state looks like Bohr atom with negatively charged particle in the core and nucleus moving along the Bohr orbit. At $2 < a < \infty$ the bound states look like Thomson atoms, in which the body of nucleus oscillates around the heavy negatively charged particle (see e.g.,~\cite{4}). 
 
In the case of $O$He $Z=2$, $Z_o=2$ and $a = Z Z_o \alpha A m_p R \le 1$, which proves its Bohr-atom-like structure~\cite{4,I,KK}.
For point-like charge distribution in helium nucleus the $O$He binding energy is given by
\begin{equation}
    E_b=\dfrac{1}{2} Z^2 Z_o^2 \alpha^2 A m_p
\end{equation}
and the radius of Bohr orbit in this ``atom''
\cite{4,I,KK} is \beq
r_{o} = \dfrac{1}{Z_{o} Z_{He} \alpha 4 m_p} = 2 \cdot 10^{-13}\cm, 
\label{ro}\eeq
being of the order of and even a bit smaller than the size of He nucleus. Therefore non-point-like charge distribution in He leads to a significant correction to the \emph{O}He binding energy.

For large nuclei or large particle charge, the system looks like Thomson atom with the particle inside the nuclear droplet. The binding
energy can be estimated in this case with the use of harmonic
oscillator approximation ~\cite{4,Cahn,Pospelov,Kohri}
\begin{equation}
    E_b=\dfrac{3}{2}(\dfrac{Z Z_o \alpha}{R}-\dfrac{1}{R}(\dfrac{Z Z_o \alpha}{A m_p R})^{1/2}).
\label{potosc}
\end{equation}

In the approximation $R_{He} \approx r_o$ one can easily find from the Equation (\ref{potosc}) that binding energy of He with \emph{X}-particle with charge $Z_o$ is given by
\begin{equation}
    E_{He}=2.4 \MeV (1-\dfrac{1}{Z_o^{1/2}}) Z_o.
\label{enx}
\end{equation}

It gives $E_{He}=4.8 \MeV$ for $X$-berillium, 8.6 MeV for $X$-carbon and 12.8 MeV for $X$-oxygen. 

$X$-nuclearites look similar to $O$-nuclearites---neutral bound states of heavy nuclei and multiple $O^{--}$ particles, compensating nuclear charge~\cite{voskr}. However, $X$-nuclearites consist of a single multiple charged lepton-like particle bound with the corresponding number of helium nuclei and hence their structure needs special study.

\subsection{Models of Stable Multiple Charged Particles}%\unskip
\subsubsection{\label{4generation} Double Cherged Stable Particles of Fourth Generation}

The existence of the fourth sequential generation can follow from heterotic string phenomenology. Its quarks and leptons can possess a new conserved charge ~\cite{4,I}. Conservation of this charge can provide stability of the lightest quark of the 4th generation~\cite{4,I}. If it is the $U$-quark, sphaleron transitions in the early Universe can establish excess of $\bar U$ antiquarks at the observed baryon asymmetry. Then  $(\bar U \bar U \bar U)$ with the charge $-$2 can be formed. It binds with $^4He$ in atom-like state of O-helium~\cite{I}. Origin of $X$-particles with larger charges seem highly unprobable in this model.
 
As we discussed above in Section \ref{SSQ} the experimental data puts constraints on possible deviation of 125 GeV Higgs boson from the predictions of the Standard model. It excludes the full strength coupling of this boson to fourth generation quarks and leptons. The suppression of these couplings implies some other nature of the mass of fourth generation  e.g., due to another heavier Higgs boson.
\subsubsection{Stable Charged Techniparticles in Walking Technicolor}

In the lack of positive result of SUSY searches at the LHC the possibilities of non-supersymmetric solutions for the problems of the Standard model become of special interest.  The minimal walking technicolor model 
%References [136] ``\cite{quentin}'' is missing. Please renumber the references so they appear in sequential numerical order.
~(WTC)~\cite{4,Gudnason:2006yj,Sannino:2004qp,Hong:2004td,Dietrich:2005jn,Dietrich:2005wk,Gudnason:2006ug} proposes the composite nature of Higgs boson. In this approach divergence of Higgs boson mass is cut by the scale of technicolor confinement. This scale also determines the scale of the electroweak symmetry breaking. Possible extensions of the minimal WTC model to improve the correspondence of this approach to the recent LHC data are discussed in~\cite{cht}.

WTC involves two techniquarks,$U$ and $D$. They transform
under the adjoint representation of a SU(2) technicolor gauge
group. A neutral techniquark--antiquark state is associated with the Higgs boson. Six bosons $UU$, $UD$, $DD$, and their 
antiparticles are technibaryons. If the technibaryon number $TB$ is conserved, the lightest technibaryon should be stable.

Electric charges of $UU$, $UD$ and $DD$ are not fixed. They are given in general by $q+1$, $q$, and $q-1$, respectively.  Here $q$ is an arbitrary real number~\cite{4,KK}. Compensation of anomalies requires in addition
technileptons $\nu'$ and $\zeta$ that are technicolor singlets with charges $(1-3q)/2$ and $(-1-3q)/2$, respectively. Conservation of technilepton number $L'$ provides stability of the lightest technilepton.

Owing to their nontrivial SU(2) electroweak charges techniparticles participate in sphaleron transitions in the early universe. Sphalerons support equilibrium relationship between $TB$,  baryon number $B$, of lepton number $L$, and  $L'$. When the rate of sphaleron transitions becomes smaller than the rate of expansion the excess of stable techniparticles is frozen out. It was shown in~\cite{KK,KK2} that there is a balance between the excess of negatively charged particles over the corresponding positively charegd particles and the observed baryon asymmetry
of the Universe. These negatively charged massive particles are bound in neutral atoms with primordial helium immediately after Big Bang nucleosynthesis. 

In the case of $q=1$ three possibilities were found for a dark atom scenario based on \mbox{WTC~\cite{4,KK,KK2}}. If $TB$ is conserved there can be excess of stable antitechnibaryons $\bar{U}\bar{U}$ with charge $-2$. If technilepton number $L'$  is conserved, the excess of stable technilepton $\zeta$ with charge $-2$ is possible.
In both cases, stable $-2$ charged particles can capture primordial $^4He^{++}$ nuclei and form $O$-helium atoms, dominating in the observed dark matter.
If both $TB$ and $L'$ are conserved a two-component techni-$O$-helium dark matter scenario is possible.

Finally,  the excessive technibaryons and technileptons can have opposite sign. Then two types of 
dark atoms, $(^4He^{++}\zeta^{--})$ and $(\zeta^{--}(U U )^{++})$, are possible. The former is nuclear interating $O$-helium, while the latter is weakly interacting and severely constrained by direct dark matter searches. Hence, WIMP-like $(\zeta^{--}(U U )^{++})$ is subdominant in this two-component scenario, while $O$-helium is the dominant component of dark matter.

In all the three cases it was shown that there are parameters of the model at which the techniparticle asymmetries have proper sign and value, explaining the $O$-helium dark matter density~\cite{KK,KK2}.

The case of multiple $-2n$ charged particles remains still unexplored for $n>1$. We have marked bold possible multiple charged candidates for stable charged constituents of $X$-nuclearites in table \ref{tab:X}. The analysis of possible structures of corresponding $X$-nuclearites and their cosmological evolution and possible impact are now under way.

\begin{table}
\caption{List of possible integer charged techniparticles. Candidates for even charged constituents of dark atoms are marked bold.}
\centering
\begin{ruledtabular}
{\begin{tabular}{ccccccccc}
\boldmath{$q$} & \boldmath{$UU(q+1)$} &  \boldmath{$UD(q)$} &  \boldmath{$DD(q-1)$} &   \boldmath{$\nu' (\dfrac{1-3q}{2} )$}& \boldmath{ $\zeta (\dfrac{-1-3q}{2} )$}\\
\hline
1 & \textbf{ 2}&  1 & 0 & $-1$ & \boldmath{$-2$} \\
3 &   \textbf{4}&  3 & \textbf{2} & \boldmath{$-4$} & $-5$\\
5 &   \textbf{6 }&   5 &  \textbf{4 }& $-7$ & \boldmath{$-8 $}\\
7 &  \textbf{ 8 }&  7 &  \textbf{ 6 }&\boldmath{ $-10$} & $-11$
\end{tabular} \label{tab:X} }
\end{ruledtabular}
\end{table}

\subsection{Effects of Hadronic Dark Matter}%\unskip
\subsubsection{Cosmology of Hadronic Dark Matter}
The considered BSM models make only a small step beyond the physics of the Standard Model and do not contain the physical basis for inflation and baryosynthesis that may provide some specific features of the cosmological scenario and mechanisms of generation of primordial density fluctuations, in particular. Therefore we assume in our cosmological scenario a standard picture of inflation and baryosynthesis with the adiabatic spectrum of density fluctuations, generated at the inflational stage. After the spectrum of fluctuations is generated, it causes density fluctuations within the cosmological horizon and their evolution depends on the matter/radiation content, equation of state and possible mechanisms of damping. The succession of steps to formation of our hadronic and hadron-like states in the early Universe needs special detailed study, but qualitatively it is similar to the evolution of tera-particles studied in~\cite{FK}.

One can divide possible forms of dark matter in hadronic and hadron-like models on two possible types. The case of new stable hadrons or composite dark matter like $O$-helium corresponds to SIMPs, while candidates without or with strongly suppressed QCD interaction are closer to WIMPs. In the latter case, the cosmological scenario should follow the main features of the Standard $\Lambda$CDM model with possible specifics related with the multicomponent WIMP-like candidates.  

In the former case, SIMP interactions with plasma support thermal equilibrium with radiation at the radiation dominance (RD) stage. The
radiation pressure acts on the plasma and then is transferred to
the $O$-helium gas. It converts $O$-helium density fluctuations in
acoustic waves, preventing their growth~\cite{4}.

At temperature $T < T_{od} \approx 1 S_3^{2/3}\keV$ SIMPs decouple from plasma and radiation, since
\cite{I,KK}  

\beq n_B \sv (m_p/m_o) t < 1. \label{eqOhe} \eeq 

Here $m_o$ is the
mass of the SIMP particle and we denote $S_3= m_o/(1 \TeV)$. In the Equation (\ref{eqOhe}) $v = \sqrt{2T/m_p}$ is the baryon thermal velocity. with the use of the analogy with $O$He case we took according to~\cite{4,I,KK}\beq \sigma
\approx \sigma_{o} \sim \pi r_{o}^2 \approx
10^{-25}\cm^2\label{sigOHe}, \eeq where $r_o$ is given by Equation(\ref{ro}). Then, SIMP gas decouples from plasma and plays the role of dark matter in formation of the large scale structure (LSS). At
 $t \sim 10^{12}\s$ corresponding to  $T
\le T_{RM} \approx 1 \eV$, SIMPs start to dominate in the Universe,
triggering the LSS formation. The details of the corresponding dark
matter scenario are determined by the nature
of SIMPs and need special study. Qualitatively, conversion in sound waves leads to suppression on the corresponding scales and the spectrum acquires the features of warmer than cold dark matter scenario~\cite{4,I,KK}. Decoupled from baryonic matter SIMP gas doesn't follow formation of baryonic objects, forming dark matter halos of galaxies.

In spite of strong (hadronic) cross section SIMP gas is collisionless on the scale of galaxies, since its collision timescale is much larger than the age of the Universe. The baryonic matter is transparent for SIMPs at large scales. Indeed, $n\sigma R = 8 \cdot 10^{-5} \ll 1$ in a galaxy with mass $M = 10^{10}M_\odot$ and radius $R = 10^{23}$ cm. Here $n = M/4\pi R^3$ and $\sigma = 2 \cdot 10^{-25}$ cm$^2$ is taken as the geometrical cross section for SIMP collisions with baryons. Therefore, SIMPs should not follow baryonic matter in formation of the baryonic objects. SIMPs can be captured only by sufficiently dense matter proto-object clouds and objects, like planets and stars (see~\cite{voskr}). 

\subsubsection{Probes for Hadronic Dark Matter}
In the charge symmetric case, SIMP  collisions can lead to indirect effects of their annihilation, like in the case of WIMP annihilation first considered in~\cite{ZKKC}, and contribute by its products to gamma background and cosmic rays. Effects of annihilation are not possible for asymmetric dark matter, but its inelastic collisions can produce cosmic particles and radiation. For example,  $O$He excitations in such collisions can result in pair production in the course of de-excitation and the estimated emission in positron annihilation line can explain the excess, observed by INTEGRAL in the galactic bulge~\cite{CKWahe}. The realistic estimation of the density of dark matter in the center of galaxy makes such explanation  possible for $O^{--}$ mass near 1.25 TeV~\cite{front}.

In the two-component dark atom model, based on the walking technicolor, a subdominant WIMP-like component $UU\zeta$ is present, with metstable technibaryon $UU$, having charge +2. Decays of this technibaryon to the same sign (positive) lepton pairs
can explain excess of high energy cosmic positrons observed by PAMELA and AMS02~\cite{AHEP}. However, any source of positrons inevitably is also the source of gamma radiation. Therefore the observed level of gamma background puts upper limit on the mass of $UU$, not exceeding 1 TeV~\cite{front}.

These upper limits on the mass of stable double charged particles challenges their search at the LHC (see ~\cite{front,Bled2017} for review and references).

Owing to their hadronic interaction, SIMP particles are captured by the Earth and slowed down in the terrestrial matter (see e.g.,~\cite{4}). After thermalization they drift towards the center of the
Earth with~velocity \beq V = \dfrac{g}{n_A \sigma_{tr} v} \approx 200\, S_3
A_{med}^{-1/6} \dfrac{1 \g/\cm^3}{\rho} \cm/\s \label{dif}\eeq 

Here $A_{med} \sim 30$ is the average
atomic weight in terrestrial surface matter. In the Equation(\ref{dif}) $$n_A=\dfrac{\rho}{A_{med} m_p}=6  \times  10^{23}/A_{med} \dfrac{\rho}{1 \g/\cm^3}$$
is the number density of terrestrial atomic nuclei and the transport geometrical cross section of collisions on matter nuclei with radius $R$, given by Equation(\ref{rA}), is given by 
$$\sigma_{tr}= \pi R^2 \dfrac{A_{med} m_p}{m_o}.$$   

We denote by $m_o$ the
mass of the SIMP particle, $S_3= m_o/(1 \TeV)$, $v = \sqrt{2T/A_{med}m_p}$ as the thermal velocity
of matter nuclei and $g=980~ \cm/\s^2$.

At a depth $L$ below the Earth's surface, the drift timescale is
$t_{dr} \sim L/V$. Here $V$ is the drift velocity given by Equation(\ref{dif}). The incoming flux changes due to the orbital motion of the Earth. It should lead to the corresponding change in the
equilibrium underground concentration of SIMPs. At the depth $L \sim 10^5\cm$ the timescale of this change of SIMP concentration is given by
$$t_{dr} \approx 5 \cdot 10^2 A_{med}^{1/6} \dfrac{\rho}{1 \g/\cm^3}S_3^{-1}\s.$$

Thermalized due their elastic collisions with matter, SIMPs are too slow to cause in the underground detectors any significant effect of nuclear recoil, on which the strategy of direct WIMP searches is based. However, a specific type of inelastic processes, combined with annual modulation of SIMP concentration can explain positive results of DAMA/NaI and DAMA/LIBRA \mbox{experiments~\cite{DAMA,DAMA-review,Bernabei:2008yi,DAMARev, DAMA2018}} in their apparent contradiction with negative results of other \mbox{experiments~\cite{CDMS, CDMS2,CDMS3,xenon,lux}}. 

In the case of $O$He such explanation was based on the existence of its 3 keV bound state with sodium nuclei~\cite{DMRev,4,Levels1}. Annual modulations in transitions to this state can explain positive results of the DAMA experiments. The rate of OHe radiative capture by a nucleus in a medium with
temperature $T$ is determined by electric dipole transition and given by~\cite{4,Levels1}
\begin{equation}
    \sigma v=\dfrac{f \pi \alpha}{m_p^2} \dfrac{3}{\sqrt{2}} (\dfrac{Z}{A})^2 \dfrac{T}{\sqrt{Am_pE}}.
    \label{radcap}
\end{equation}

Here $A$ and $Z$ are atomic number and charge of nucleus, $E$ is the energy level. The factor $f = 1.4 10^{-3}$ accounts for violation of isospin symmetry in this electric dipole transition since He nucleus in $O$He is scalar and isoscalar.
Since the rate of the $O$He radiative capture is proportional to the temperature (or to the product of mass of nucleus and square of relative velocity in the non-equilibrium case)  the effect of such capture shpould be suppressed in cryogenic detectors. On the other hand, the existence of a low energy bound state was found in~\cite{Levels1} only for intermediate mass nuclei and excluded for heavy nuclei, like xenon. It can explain the absence of the signal in such experiments as XENON100~\cite{xenon} or LUX~\cite{lux}. The confirmation of these results should follow from complete and self-consistent quantum mechanical description of $O$He interaction with nuclei, which still remains an open problem for the dark atom scenario.

\subsection{Open Problems of Hadronic Dark Matter}

In spite of uncertainty in the description of the interaction with matter of hypercolor motivated dark matter candidates, they are most probably similar to WIMPs and thus can hardly resolve the puzzles of direct dark matter searches.

One should note that in the simplest case hadronic dark matter can also  hardly provide solution for these puzzles.  Slowed down in the terrestrial matter SIMP ellastic collisions cannot cause significant nuclear recoil in the underground detectors, while inelastic nuclear processes should lead to energy release in the MeV range and cannot provide explanation for the signal detected by DAMA experiments in a few keV range. This puzzle may be resolved by some specifics of structure of SIMPs and their interaction with nuclei. 

Such specifics was proposed in the dark atom model and the solution of a low energy bound state of the $O$He-nucleus system was found.
This solution was based on the existence of a dipole repulsive barrier that arises due to $O$He polarization by the nuclear attraction of the approaching nucleus and provides a shallow potential well, in which the low energy level is possible for intermediate mass nuclei.
However, the main open problem of the dark atom scenario is the lack of the correct quantum mechanical treatment of this feature of $O$He nuclear interaction. In the essence, this difficulty lies in the necessity to take into account simultaneous effect of nuclear attraction and Coulomb repulsion in the absence of the usual simplifying conditions of the atomic physics (smallness of the ratio of the core and the shell as well as the possibility of perturbative treatment of the electromagnetic interaction of the electronic shell). In any case, strongly interacting dark matter cannot cross the matter from the opposite side of the Earth and it should inevitably lead to diurinal modulation of the $O$He concentration in the underground detector and the corresponding events. The role of this modulation needs special study for the conditions of the DAMA/NaI and DAMA/LIBRA experiments.

The lack of correct quantum mechanical treatment of $O$He nuclear physics also leaves open the question on the dominance of elastic collisions with the matter, on which $O$He scenario is based and on the possible role of inelastic processes in this scenario. Indeed, screening the electric charge of the $\alpha$-particle, $O$He interactions after Big Bang nucleosynthesis can play a catalyzing role in production of primordial heavy elements, or influence stellar nucleosynthesis. Therefore the lack of developed $O$He nuclear physics  prevents detailed analysis of possible role of $O$He in nucleosynthesis, stellar evolution and other astrophysical processes, as well as elaboration of the complete dark atom scenario.

The attractive feature of the dark atom model is the  possibility to explain the excess of positron annihilation line emission, observed by INTEGRAL, and the excess of high energy fraction of cosmic positrons, detected by PAMELA and AMS02. These explanations are possible for double charged particles with the mass below 1.3 TeV, challenging the probe of the existence of its even-charged constituents in the direct searches at the LHC~\cite{Bled2017}, in which effects of two-photon annihilation of bound multiple charged particles should be also taken into account~\cite{multiple}. 

\section{Composite Dark Matter in the Context of Cosmo--Particle Physics}\label{Discussion}

Observational cosmology offers strong evidence in favor of a new physics that challenges its discovery and thorough investigation.
Cosmo--particle physics~\cite{ADS,MKH,book,newBook} elaborates methods to explore new forms of matter and their physical
properties. Physics of dark matter plays an important role in this process and we make here a small step in the exploration of possible forms of this new physics.

%\section{ as the solution of Ouroboros puzzle}
 %\label{Ouroboros}
 
 %\section{Conclusions}
 
We consider some scenarios of SM extensions with new strongly interacting heavy particles  which are suggested as DM candidates (new stable hadrons, hyperhadrons and nuclear interacting dark atoms). It makes a minor step beyond the physics of the Standard model and doesn’t provide mechanisms for inflation and baryosynthesis. But even such a modest step provides many  new interesting physical and astrophysical phenomena.  
 
Within the framework of the hadronic scenario, we analyzed the simplest extensions of the SM quark sector with new heavy quark in fundamental representation of color $\text{SU}_\text{C}(3)$ group. It was shown that the prediction of new heavy
hadrons, which consist of new and standard quarks, does not contradict to cosmological constraints. 

 An appearance of new heavy quarks resulted from some additional vacuum symmetry breaking at a scale which is much larger than EW scale. Namely, in the mirror model (it is one of possible scenarios with new heavy fermions), there should be an extra Higgs doublet with v.e.v. $\sim$(10$^6$--10$^{10}$)~GeV to provide masses for new fermions~\cite{10}. Analogously, an origin of heavy singlet quark as a consequence of the chain of transitions like $E(6)\to SO(10)\to SU(5)$ also stems from an additional Higgs doublet and corresponding symmetry breaking at a high scale. We suppose that both of these variants of heavy fermion emergence take place at the end of the inflation stage (or~after it) due to some first Higgs transition. Then, massive non-stable states decay contributing to the quark-gluon plasma, while stable massive quarks go on interacting with photons. Later, after the second (electroweak) Higgs transition, remaining ``light'' quarks become massive, and (at the hadronic epoch) hadrons can be formed as bound states of the quarks. New ``heavy-light'' mesons also produced after the EW symmetry breaking. At the same time, we can just appeal to some special unknown dynamics which should break the symmetry at some higher scale. It may cause nontrivial features of cosmological scenario, which deserve further analysis.

 There also arises a question: can the heavy dark matter occurrence substantially modify the distribution of the energy density produced by the acoustic waves at the inflation stage? As we suppose, the most intensive inflation sound waves had passed through the plasma before the first Higgs transition, an emergence of massive particles after the passing of the waves (when the inflation is finished) would only insufficiently change the density in already-formed areas with high energy density. In other words, this primary DM does not qualitatively change the whole pattern of the density distribution imprinting in CMB.

Here we show that the schemes with extensions of quark sector are in agreement with the precision electro-weak constraints on new physics effects. Using an effective Lagrangian of low-energy hadron interaction, we get the asymptotics of potential of interaction.
In principle, both lighter mesons $M = (Qq)$ and fermions $B = (Qqq)$ can be stable.However, fermionic states can burn out in collisions with nucleons and new heavy mesons.  

The main argument in favor of choosing mesons (two-quark pseudoscalars) as the DM candidates is that they have a repulsive potential for interactions with nucleons. This is due to the absence of one-pion exchanges, since the corresponding vertex is forbidden by parity conservation. Such a vertex for new fermions is not forbidden, and, at large distances, the potential of their interaction can be attractive. As a result, the new fermions can form bound states with nucleons, but not mesons (at low energies). In other words, at long distances  between new heavy hadrons and also between them and nucleons a repulsive forces arise, and a potential barrier prevents the formation of bound states of new and standard hadrons at low energies. From the experimental limits on anomalous hydrogen and helium we can conclude that $M^0$ and $\bar{M}^0$ abundances are not equal, and this asymmetry is opposite to ordinary baryon asymmetry. An alternative scenario (symmetrical abundance of new quarks) can be permissible when new hadrons are superheavy (with mass $M > 10$ TeV).

As it was shown here, at the galaxy scale even hadronic DM behaves as collision-less gas. Therefore, both bosons and fermions should be distributed in the galactic halo. The difference between them can appear only after the DM capture by stars, when the difference in statistics can be important for a DM core inside the star. Indeed, type of the DM should influence on the formation and evolution of stars. Certainly, for scalars (with the repulsive interaction) and fermions (with an attractive potential at low energies), these processes are different. Discussion of these interesting problems is, however, beyond the scope of this paper.

Taking into account experimental data on masses of standard heavy-light mesons we evaluated the mass-splitting of charged $M^-$ and neutral $M^0$ components and calculated width of charged meson which occurs large, so the charge component is long-lived, $\tau\gg 1\,s$. Starting from the relic concentration of dark matter and the expression for annihilation cross-section, the mass of the dark matter candidate was determined. The estimation of mass without SGS enhancement gives the value of mass near 20 TeV and an account of this effect increases this value up to $10^2$ TeV. These estimations are in agreement with the evaluations of mass in the scenarios with baryonic DM,
which are considered in literature. Thus, in the LHC experiments, superheavy new hadrons cannot be produced and directly detected in the nearest future. Moreover, it is difficult to observe superheavy hadrons when searching and studying an anomalous hydrogen and helium. However, as it was noted early, charged hadron $M^-$ having a large lifetime can be directly detected in the process of $M^0 N$ scattering off energetic nucleons. In the calculation of annihilation cross-section we take into account some peculiarities of SGS effect. We note also that annihilation cross-section was considered here at the level of sub-processes. It means, to analyze features of the hadronic DM in more detail we need to clarify the annihilation mechanism.

We can conclude that the extensions of the SM with additional heavy fermions and vector-like interactions are perspective both from the theoretical point of view (they demonstrate an interesting structure of dynamical symmetry group and a wide spectrum of states) and an area for the checking of model predictions in collider experiments and in astrophysics. 

We have also considered some particular representatives in the class of SM extensions with an additional hypercolor gauge group that confines a set of new vector-like fermions, H-quarks. If the hypercolor group is chosen as symplectic one, the global symmetry of the model is SU(2$n_F$) ($n_F$ is a number of H-quark flavors), which is larger than the chiral group and predicts consequently a spectrum of H-hadrons that contains not only heavy analogues of QCD hadrons, but also new states such as heavy diquarks (H-baryons). We have considered the cases of two and three H-flavors.

The analysis of the oblique parameters showed that in order to comply with the restrictions of precision data of the SM, it is necessary to fulfil certain conditions imposed on this type of extensions. For example, there are lower bounds on the masses of new particles arising in the framework of the model, and the mixing parameter of scalars, Higgs boson and H-sigma. The study of the symmetry properties led to the conclusion that two stable neutral objects exist in the SU(4) scenario (a more general SU(6) version with partially composite Higgs is being studied). Thus, the~model predicts a two-component dark matter structure. The complexity of the DM has been discussed \mbox{repeatedly~\cite{DMRev,4,KK2,Multi_1,Multi_2,Two_Comp,Two_Comp_2,Multi_N}}, starting from the early works~\cite{UDM,UDM1,MCDM} but in the case of the vector-like model, the emergence of two components is a consequence of the symmetry of the model, and not an artificial assumption to explain some of the observed features in the measured spectra of cosmic photon or lepton fluxes. Such inclusion of various DM components in the unique theoretical framework can be a natural consequence of the realistic extensions of the SM symmetry, as it was shown in the gauge models of broken quark-lepton family symmetry~\cite{Berezhiani1,Sakharov1}.

Hoping for the detection of specific signals from new H-hadrons at the collider (when extending the energy interval and improving statistics), it is necessary also to deepen an analysis of the DM particle effects in astrophysics. It is especially important to study the channels of interaction of the neutral component of the H-diquark with ordinary fermions---these reactions define significantly of the observed features of leptonic and photonic spectra measured by space telescopes.

The started analysis of the DM scenarios in the framework of the SU(6) extension promises to be very interesting since the dark matter in this case may seem to contain another H-hadron component. Of course, in hyper-color models, the hyper-interaction itself is not studied yet, but its scale is noticeably higher than the achieved energies at the collider, and the corresponding loop contributions of hyperquarks, hypergluons and heavy H-hadrons are too small to noticeably affect the spectra observed in astrophysics. It seems that an analysis of high-energy cosmic rays scattering off the DM, the consideration of the annihilation of DM particles in various clump models, have good prospects. Analysis of photon and lepton signals will allow not only to assess the validity and possibilities of the proposed options for the SM extension, but also to understand how the DM multicomponent structure arising in them manifests itself in the observed data and gives an information on the DM distribution in the galaxy.

The dark atom scenario is based on the minimal extension of the SM content, involving only hypothetical stable even charged particles and reducing most of observable DM effects to the properties of the helium shell of $O$He and its nuclear interactions. Such effects can provide nontrivial solutions for the puzzles of direct and indirect dark matter searches and it looks like this model can be made fully predictible on the basis of the known physics. However, the nontrivial features of $O$He interaction with nuclei still leave an open problem a self-consistent quantitative analysis of the corresponding~scenario.

To conclude, we have discussed a set of nontrivial dark matter candidates that follow from the BSM models, involving QCD color, hypercolor and technicolor physics. These predictions provide interesting combinations of collider, non-collider, astrophysical and cosmological signatures that can lead to thorough investigation of these models of new physics by the methods of cosmo--particle physics. We cannot expect that our models can give answers to all the problems of the physical basis of the Universe, but the observational fact that we live in the Universe full of unknowm forms of matter and energy stimulates our efforts to approach the mystery of their puzzling nature.

\vspace{6pt}
%%%%%%%%%%%%%%%%%%%%%%%%%%%%%%%%%%%%%%%%%%

\acknowledgments{Authors are sincerely grateful to M. Bezuglov for fruitful cooperation in scientific research and discussion of the results. The work was supported by a grant of the Russian Science Foundation (Project No-18-12-00213).}

\appendix
\section{\label{app:gens}Algebras su(6) and sp(6)}

The generators $\Lambda_\alpha$, $\alpha=1,\dots,\,35$ of SU(6) satisfy the following relations:
\begin{gather}
    \Lambda_\alpha^\dagger = \Lambda_\alpha, \qquad \Tr \Lambda_\alpha = 0, \qquad \Tr \Lambda_\alpha \Lambda_\beta = \dfrac12 \delta_{\alpha\beta}.
\end{gather}

It is convenient for us to separate two subsets of the generators. The first one forms a sub-algebra $\text{sp}(6) \subset \text{su}(6)$---it includes matrices $\Sigma_{\dot\alpha}$, $\dot\alpha=1,\dots,\,21$ satisfying a relation 
\begin{gather}
    \Sigma_{\dot\alpha}^T M_0 + M_0 \Sigma_{\dot\alpha} =0,
\end{gather}
where the antisymmetric matrix $M_0$ is chosen as in Equation~\eqref{eq:LTQ}.
%\begin{gather}
%    M_0 = \begin{pmatrix} 0 & \varepsilon & 0 \\ \varepsilon & 0 & 0 \\ 0 & 0 & \varepsilon \end{pmatrix}.
%\end{gather}

We choose these generators as follows:
\begin{gather*}
    \Sigma_{a} = \dfrac{1}{2\sqrt2} \begin{pmatrix} 0 & \tau_a & 0 \\ \tau_a & 0 & 0 \\ 0 & 0 & 0 \end{pmatrix},
    \quad
    \Sigma_{3+a} = \dfrac{i}{2\sqrt2} \begin{pmatrix} 0 & \tau_a & 0 \\ -\tau_a & 0 & 0 \\ 0 & 0 & 0 \end{pmatrix},
    \quad
    \Sigma_{6+a} = \dfrac{1}{2} \begin{pmatrix} 0 & 0 & 0 \\ 0 & 0 & 0 \\ 0 & 0 & \tau_a \end{pmatrix},
\\
    \Sigma_{9+a} = \dfrac{1}{2\sqrt2} \begin{pmatrix} \tau_a & 0 & 0 \\ 0 & \tau_a & 0 \\ 0 & 0 & 0 \end{pmatrix},
    \quad
    \Sigma_{12+a} = \dfrac{1}{4} \begin{pmatrix} 0 & 0 & \tau_a \\ 0 & 0 & \tau_a \\ \tau_a & \tau_a & 0 \end{pmatrix},
    \quad
    \Sigma_{15+a} = \dfrac{i}{4} \begin{pmatrix} 0 & 0 & \tau_a \\ 0 & 0 & -\tau_a \\ -\tau_a & \tau_a & 0 \end{pmatrix},
\\
    \Sigma_{19} = \dfrac{1}{2\sqrt2} \begin{pmatrix} 1 & 0 & 0 \\ 0 & -1 & 0 \\ 0 & 0 & 0 \end{pmatrix},
    \quad
    \Sigma_{20} = \dfrac{1}{4} \begin{pmatrix} 0 & 0 & 1 \\ 0 & 0 & -1 \\ 1 & -1 & 0 \end{pmatrix},
    \quad
    \Sigma_{21} = \dfrac{i}{4} \begin{pmatrix} 0 & 0 & 1 \\ 0 & 0 &1 \\ -1 & -1 & 0 \end{pmatrix},
\end{gather*}
where $\tau_a$, $a=1,\,2,\,3$ are the Pauli matrices.

The remainder of the generators, $\beta_\alpha$, $\alpha=1,\dots,\,14$, belong to the coset SU(6)/Sp(6) and \mbox{satisfy a relation}
\begin{gather}
    \beta_\alpha^T M_0 = M_0 \beta_\alpha.
\end{gather}

These can be defined as follows:
\begin{gather*}
    \beta_1 = \dfrac{1}{2\sqrt2} \begin{pmatrix} 0 & 1 & 0 \\ 1 & 0 & 0 \\ 0 & 0 & 0 \end{pmatrix},
    \qquad
    \beta_2 = \dfrac{i}{2\sqrt2} \begin{pmatrix} 0 & -1 & 0 \\ 1 & 0 & 0 \\ 0 & 0 & 0 \end{pmatrix},
    \qquad
    \beta_{2+a} = \dfrac{1}{2\sqrt2} \begin{pmatrix} \tau_a & 0 & 0 \\ 0 & -\tau_a & 0 \\ 0 & 0 & 0 \end{pmatrix},
\end{gather*}
\begin{gather*}
    \beta_{6} = \dfrac{1}{2\sqrt6} \begin{pmatrix} 1 & 0 & 0 \\ 0 & 1 & 0 \\ 0 & 0 & -2 \end{pmatrix},
    \qquad
    \beta_{6+a} = \dfrac{1}{4} \begin{pmatrix} 0 & 0 & \tau_a \\ 0 & 0 & -\tau_a \\ \tau_a & -\tau_a & 0 \end{pmatrix},
    \qquad
    \beta_{10} = \dfrac{1}{4} \begin{pmatrix} 0 & 0 & 1 \\ 0 & 0 & 1 \\ 1 & 1 & 0 \end{pmatrix},
\\  \beta_{10+a} = \dfrac{i}{4} \begin{pmatrix} 0 & 0 & \tau_a \\ 0 & 0 & \tau_a \\ -\tau_a & -\tau_a & 0 \end{pmatrix},
    \qquad
    \beta_{14} = \dfrac{i}{4} \begin{pmatrix} 0 & 0 & 1 \\ 0 & 0 & -1 \\ -1 & 1 & 0 \end{pmatrix}.
\end{gather*}

%=============================================================================
%=============================================================================
%=============================================================================

\end{document}